\documentclass[a4paper,fleqn]{cas-sc}

\usepackage[numbers]{natbib}
\usepackage{amsmath}
\usepackage{todonotes}
\usepackage{float}
\usepackage{booktabs}
\usepackage{threeparttable}
\usepackage{subcaption}
\usepackage{soul}
\usepackage{xcolor}
\usepackage[section]{placeins}

\renewcommand{\vec}[1]{\textbf{#1}}
\newcommand\norm[1]{\lVert#1\rVert}
\newcommand{\ignore}[1]{}

\def\tsc#1{\csdef{#1}{\textsc{\lowercase{#1}}\xspace}}
\tsc{WGM}
\tsc{QE}

\begin{document}

\shorttitle{Fluid-structure coupled simulation of explosion containment}    

\shortauthors{A.~Narkhede {\it et al.}}  

\title [mode = title]{
	Fluid-structure coupled simulation framework for lightweight explosion containment structures under large deformations
}



%

\author[1]{Aditya Narkhede}[orcid=0009-0004-4281-611X,
linkedin=adityanarkhede]
%
%
\ead{anarkhede@vt.edu}
%
%
%
\affiliation[1]{organization={Kevin T. Crofton Department of Aerospace and Ocean Engineering, Virginia Tech},
	city={Blacksburg},
	postcode={24061}, 
	state={Virginia},
	country={USA}}
\author[1]{Shafquat Islam}
\author[2]{Xingsheng Sun}
\author[1]{Kevin Wang}

\affiliation[2]{organization={Department of Mechanical and Aerospace Engineering, University of Kentucky},
	city={Lexington},
	postcode={40506},
	state={Kentucky},
	country={USA}}



%
%
%

	\begin{abstract}
		Lightweight, single-use explosion containment structures provide an effective solution for neutralizing rogue explosives, combining affordability with ease of transport. This paper introduces a three-stage simulation framework that captures the distinct physical processes and time scales involved in detonation, shock propagation, and large, plastic structural deformations. A working hypothesis is that as the structure becomes lighter and more flexible, its dynamic interaction with the gaseous explosion products becomes increasingly significant. Unlike previous studies that rely on empirical models to approximate pressure loads, this framework employs a partitioned procedure to couple a finite volume compressible fluid dynamics solver with a finite element structural dynamics solver. Given the rapid expansion of explosion products and the large structural deformation, the level set and embedded boundary methods are utilized to track the fluid-fluid and fluid-structure interfaces. The interfacial mass, momentum, and energy fluxes are computed by locally constructing and solving one-dimensional bi-material Riemann problems. A case study is presented involving a thin-walled steel chamber subjected to an internal explosion of $250~\text{g}$ TNT. The result shows a $30\%$ increase in the chamber volume due to plastic deformation, with its strains remaining below the fracture limit. Although the incident shock pulse carries the highest pressure, the subsequent pulses from wave reflections also contribute significantly to structural deformation. The high energy and compressibility of the explosion products lead to highly nonlinear fluid dynamics, with shock speeds varying across both space and time. Comparisons with simpler simulation methods reveal that decoupling the fluid and structural dynamics overestimates the plastic strain by $43.75\%$, while modeling the fluid dynamics as a transient pressure load fitted to the first shock pulse underestimates the plastic strain by $31.25\%$.

	\end{abstract}




	\maketitle
	
	\section{Introduction}\label{sec:introduction}
	
	Explosions are rapid exothermic chemical reactions that produce high-velocity, high-pressure gaseous products at elevated temperatures. One way to mitigate their destructive impact is by confining the explosion within a container capable of dissipating the released energy. Traditional explosion containment structures  feature thick steel walls, designed to restrict their response within the limit of elastic vibrations. While suitable for repeated use, these structures are often heavy, expensive, and difficult to transport, making them less accessible for local public safety agencies. In contrast, lightweight, compact, and low-cost containment structures, even if designed for single use, can be a more practical and attractive option. To achieve this goal, the structure may be allowed to undergo large, permanent deformation, but not fracture.
	
	In this work, we consider explosions with a supersonic reaction front, i.e., detonations. Predicting the structural response in such events requires resolving multiple physical processes with distinct time scales. The duration of chemical reaction, $t_{\text{react}}$, is determined by the quantity of explosive material and the velocity of the reaction front. The shock travel time across the containment structure, $t_{\text{shock}}$, depends on the shock speed in the gaseous explosion product. The attenuation time of structural vibration, $t_{\text{vibr}}$, depends on the structure's natural frequencies. Typically,
	\begin{equation}
		t_{\text{react}} \ll t_{\text{shock}} \ll t_{\text{vibr}}.
	\end{equation}
	Compared to external explosions that occur in an open environment, internal explosions are more challenging in that the structure is subjected to repeated impulsive loads due to wave reflections within the confined space. During this process, the structural dynamics also influences the internal fluid flow reciprocally.
	
	Previous studies have often neglected this two-way fluid-structure interaction, instead estimating pressure loads using either empirical models (e.g., Conventional Weapons program (ConWep), exponentially decaying functions) or decoupled fluid dynamics analyses. These simplifications can be justified for many traditional applications involving open structures or thick-walled containers undergoing small deformation. For example, analytical models have been developed to estimate the maximum strains within containers, assuming linear elasticity while disregarding wave reflections~\cite{baker_elastic_1960,duffey_containment_1973, duffey_detonation-induced_2002, duffey_strain_2003}. Through computational analysis, Ma {\it et al}. showed that for a thick-walled circular cylinder with open ends, the difference between coupled and decoupled pressure loads is less than $5\%$~\cite{ma_failure_2010, ma_ductile_2013}. However, even for thick-walled structures, it has been found that when reflected waves reverberate around the structure's elastic beating frequency, late-time resonance can occur. This leads to an increase in elastic strain, a phenomenon known as strain-growth~\cite{dong_interactive_2010}.
	
	We hypothesize that as the explosion containment structure becomes lighter and more flexible, its interaction with the flow of gaseous explosion products becomes increasingly significant. While reflected shock pulses could amplify deformations, an increase in container volume would reduce the energy density of the explosion products. Neglecting the fluid-structure interaction effects could lead to either overestimation or underestimation of the structural response, depending on the specific simplifications made. Therefore, the analysis and design of single-use containment structures may need to account for the dynamic two-way interaction between the containment structure and the internal fluid flow.
	
	Some researchers have applied Arbitrary Lagrangian-Eulerian (ALE) methods --- for example, the one implemented in LS-Dyna \cite{souli_ale_2000,alia_high_2006} --- to conduct coupled fluid-structural analysis of partially confined structures (e.g.,~\cite{langdon_response_2014, li_internal_2021}) and thick-walled containers (e.g.,~\cite{dong_interactive_2010, liu_dynamic_2020}). Also, Pickerd {\it et al.} simulated welded cuboid containers subjected to large deformations, focusing on stress concentrations around welded joints~\cite{pickerd_analysis_2016}. They compared results obtained with blast load estimates from ConWep \cite{army_fundamentals_1986}, quasi-static pressure loading, and ALE, showing that ConWep significantly underestimates maximum deformation. The authors attributed the discrepancies to multiple shock reflections and subsequent quasi-static internal pressures, which contributed most to the containers' peak deformations. Overall, these studies primarily focus on the structural responses, while details of the internal fluid dynamics (e.g.,~shock reflections and interactions) are often not provided. A recently developed alternative couples LS-Dyna's Conservation Element and Solution Element (CESE) solver \cite{chang_method_1995} with its finite element solver, which has been used for fluid-structure analysis of confined structures under gas mixture detonations (e.g. \cite{rokhy_tracking_2023, rokhy_investigation_2022}).
		
	
	We present a three-stage, fluid-structure coupled computational framework to simulate internal explosions within lightweight structures undergoing large, permanent deformations. The detonation reaction is modeled using Chapman-Jouguet theory~\cite{fickett_detonation_2000}, which assumes an infinitely thin reaction front propagating at a constant speed. After the reaction completes, the resulting fluid flow exhibits three key features: a shock wave advancing through quiescent air, a receding rarefaction within the explosion products, and an expanding interface between the explosion products and the air. We simulate the shock-driven fluid flow by solving the compressible inviscid Navier-Stokes equations in the Eulerian frame using a high-resolution finite volume method. The material interface is tracked implicitly using the level-set method \cite{main_enhanced_2017,zhao_simulating_2023}. Interfacial mass, momentum, and energy fluxes are computed using the FInite Volume method with Exact multi-material Riemann problems (FIVER), which naturally handles strong discontinuities in state variables (e.g., density, internal energy) and differing equations of state across the material boundary~\cite{farhat_fiver_2012, main_enhanced_2017, zhao_long_2023}.
	
	The response of the thin-walled containment structure is simulated using a nonlinear finite element method. Interactions between the fluid flow and the structure are captured through a partitioned procedure~\cite{farhat_robust_2010}. To accommodate large deflections, we use an embedded boundary method, which tracks the fluid-structure interface on a fixed, non-body-conforming fluid mesh, avoiding the need of mesh motion and re-meshing~\cite{wang_computational_2012, wang_computational_2015, ma_computational_2022}. The normal velocity is continuous across the interface, enforced by solving a 1D fluid-structure Riemann problem locally featuring a constant interface velocity obtained from the structural solver~\cite{wang_computational_2012,ma_computational_2022}. This framework has been implemented in open-source codes~\cite{aerof,m2c,aeros}, and validated for several multiphase flow and fluid-structure interaction problems. These include underwater explosions and implosions~\cite{farhat_dynamic_2013, ma_computational_2022},  cavitation~\cite{cao_shock_2021,zhao_long_2023}, hypervelocity impacts~\cite{islam_fluid_2023,islam_plasma_2023}, and shockwave and laser lithotripsy~\cite{wang_multiphase_2017, cao_shock_2019, zhao_simulating_2023}. 
	
	The remainder of the paper is organized as follows. Section~\ref{sec:gov_eq} introduces the physical models and computational methods used in the simulation framework. The framework is verified through three numerical experiments, addressing detonation front propagation, shock-dominated flow, and blast-loaded structures. These test cases are presented in Section~\ref{sec:verification}. In Section~\ref{sec:results}, we present a case study of a thin-walled containment structure subjected to a $250~\text{g}$ TNT explosion. The dynamics of pressure, velocity, and total energy within the containment structure are analyzed. In Section~\ref{sec:fluid_estimates}, we present a comparison between the coupled simulation and two decoupled methods. The first method treats the containment structure as a rigid wall when simulating the internal fluid flow, while the second uses the Friedlander empirical model~\cite{friedlander_diffraction_1946} to estimate the pressure load. Finally, concluding remarks are provided in Section~\ref{sec:conclusion}.

	\section{Computational model}\label{sec:gov_eq}
	
\begin{figure}[!htb]
		\centering\includegraphics[width=150mm]{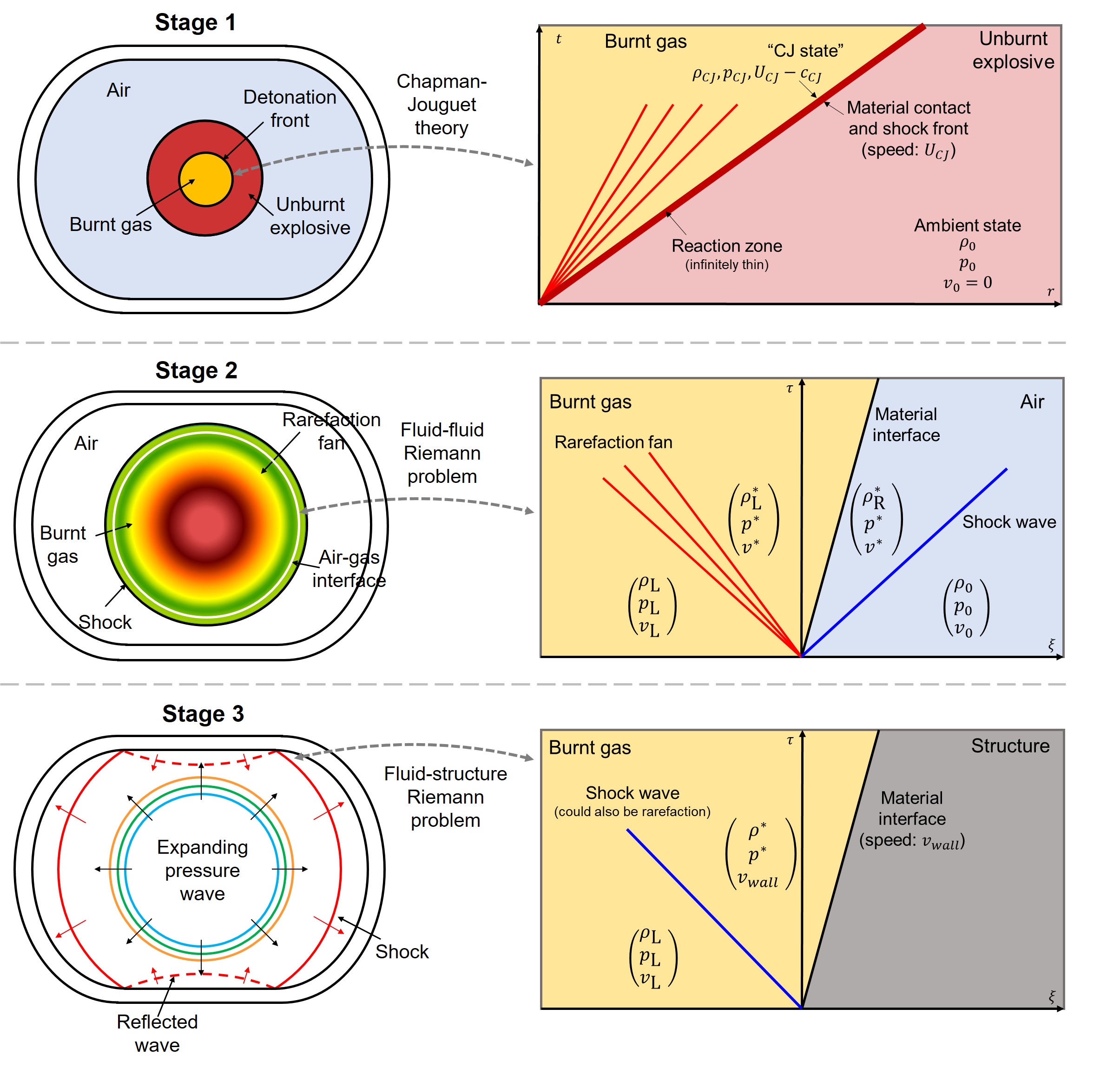}
		\caption{A three-stage simulation procedure: The global models (left) and local models applied at material interfaces (right).}
		\label{fig:model_problem}
	\end{figure}

	An internal explosion typically involves four key physical processes: (1) detonation of the explosive material, (2) propagation of shock waves and the expansion of gaseous explosion products (``burnt gas'') within the containment structure, (3) interaction of the fluid flow with the inner wall of the structure, and (4) the deformation and potential failure of the structure. We present a three-stage simulation procedure, as illustrated in Figure~\ref{fig:model_problem}. In the first stage, we model the detonation process based on Chapman-Jouguet theory, assuming a spherical explosive charge ignited at its center. The second stage involves simulating the fluid dynamics within the containment structure until the detonation-induced shock wave reaches the structure. The mixing between the burnt gas and the ambient air occurs over a longer time scale. Therefore, at this stage we assume that they remain separated by a sharp interface. The assumption of spherical symmetry is still valid, and leveraged to reduce the computational cost. In the third stage, the fluid density, velocity, and pressure fields from the previous stage are employed to initialize a three-dimensional (3D) fluid-structure interaction simulation, which predicts the dynamic structural response of the containment system.

	\subsection{Stage 1: The detonation process}\label{sec:detonation}
	
	Assuming spherical symmetry, the spatial domain of the explosive charge can be defined as $0\leq r \leq r_0$, where $r_0$ denotes the radius of the charge. If the density and mass of the charge are given by $\rho_0$ and $m_0$, respectively, then
	\begin{equation}
		r_0 = \Big(\dfrac{3m_0}{4\pi\rho_0}\Big)^{\frac{1}{3}}.
	\end{equation}
	
	To describe the chemical reaction of explosion, we adopt the Chapman-Jouguet (CJ) theory, which assumes an infinitely thin reaction zone (Figure~\ref{fig:model_problem}). Thus, the burnt gas immediately behind the detonation front can be related to the unreacted explosive through the Rankine-Hugoniot jump conditions, i.e.,
	\begin{align}
		U_{\text{CJ}} &= \dfrac{\rho_{\text{CJ}}}{\rho_\text{0}}c_{\text{CJ}},\nonumber\\
		p_0+\rho_0 U_{\text{CJ}}^2&=p_{\text{CJ}}+\rho_{\text{CJ}}c_{\text{CJ}}^2,\label{eq:cj_equations}\\
		q_\text{det}+\frac{p_0}{\rho_0}+\dfrac{U_{\text{CJ}}^2}{2}&=e_{\text{CJ}}+\frac{p_{\text{CJ}}}{\rho_\text{CJ}}+\frac{c_{\text{CJ}}^2}{2},\nonumber    
	\end{align}
	where $p_0$ denotes the pressure of the explosive charge, and $q_\text{det}$ the heat of detonation for the explosive material. $U_\text{CJ}$ denotes the constant propagation speed of the detonation front. $\rho_\text{CJ}$ and $p_\text{CJ}$ are the density and pressure of the burnt gas immediately behind the detonation front. $e_\text{CJ}$ is the internal energy of the gas. $c_\text{CJ}$ is the speed of sound calculated from $\rho_\text{CJ}$ and $p_\text{CJ}$.

	
	We model the thermodynamics of the burnt gas by the Jones-Wilkins-Lee (JWL) equation of state (EOS):
	\begin{equation}
		\label{eq:jwl}
		p=A_1\left( 1-\frac{\omega\rho}{R_1\rho_0}\right)\exp\left[ -\frac{R_1\rho_0}{\rho}\right]+A_2\left( 1-\frac{\omega\rho}{R_2\rho_0}\right)\exp\left[ -\frac{R_2\rho_0}{\rho}\right]+\omega\rho e,
	\end{equation}
	where $p$, $\rho$, and $e$ are the fluid pressure, density, and internal energy, respectively. $A_1$, $A_2$, $R_1$, $R_2$, and $\omega$ denote the material-specific parameters. The speed of sound, $c$, is given by
	\begin{equation}
		\label{eq:jwl_soundspeed}
		c = 
		\sqrt{\dfrac{1}{\rho}\Big(A_1\Big(\dfrac{R_1\rho_0}{\rho} +\omega-1\Big)\text{exp}\Big[\dfrac{-R_1\rho_0}{\rho}\Big]+ A_2\Big(\dfrac{R_2\rho_0}{\rho}+\omega-1\Big)\text{exp}\Big[\dfrac{-R_2\rho_0}{\rho}\Big]+(1+\omega)p\Big)}.
	\end{equation}
	
	The explosive material is assumed to be trinitrotoluene (TNT), with parameter values given in Table~\ref{tab:eos_parameters}. Solving Equation~\eqref{eq:cj_equations} with these parameters values yields $\rho_{\text{CJ}} = 2.2260\times10^{-3}~\text{g}/\text{mm}^3$, $U_{\text{CJ}} = 6.9154\times10^6~\text{mm}/\text{s}$, and $p_{\text{CJ}} = 2.0872\times10^{10}~\text{Pa}$.
	
	We model the burnt gas as a compressible inviscid fluid, as the detonation creates a high-speed, high-pressure flow. The analysis starts at the time of ignition. Therefore, the spatial domain of burnt gas is initially empty, and it expands radially at speed $U_\text{CJ}$. Leveraging spherical symmetry, the governing Euler equations and boundary conditions are given by
	\begin{equation}
		\frac{\partial }{\partial t}
		\begin{bmatrix}
			\rho\\
			\rho u_\text{r}\\
			E
		\end{bmatrix} + \frac{\partial}{\partial r}
		\begin{bmatrix}
			\rho u_\text{r} \\
			\rho u_\text{r}^2 + p\\
			\left(E+p\right)u_\text{r}
		\end{bmatrix}=-\frac{2}{r}
		\begin{bmatrix}
			\rho u_\text{r} \\
			\rho u_\text{r}^2\\
			\left(E+p\right)u_\text{r}
		\end{bmatrix},
		\quad 0 < r < U_{\text{CJ}}t,\quad 0 < t < \dfrac{r_0}{U_{\text{CJ}}},
		\label{eq:sp_euler}
	\end{equation}
	\begin{align}
		u_r(0, t)&=0, \color{red}\\
		\rho(U_{\text{CJ}}t, t)&=\rho_{\text{CJ}},\\
		u_r(U_{\text{CJ}}t, t)&=U_{\text{CJ}}-c_{\text{CJ}},\\
		p(U_{\text{CJ}}t, t)&=p_{\text{CJ}},
	\end{align}
	where $\rho$, $u_\text{r}$, $p$, and $E=\rho e+\dfrac{1}{2}\rho u_\text{r}^2$ denote the fluid density, radial velocity, pressure, and total energy per unit volume, respectively. This boundary value problem can be solved in different ways. One approach is to exploit the solution's self-similar structure, reducing the governing partial differential equations (PDEs) to two ordinary differential equations (ODEs) for $u_r$ and $c$. Taylor presented a numerical solution to these ODEs using the perfect gas equation of state (EOS) for the burnt gas \cite{taylor_dynamics_1950}. Their method can be adapted to the JWL EOS. Another approach, which is adopted in this work, is to numerically solve the governing PDEs directly.
	
	We use a standard high-resolution finite volume method to discretize~\eqref{eq:sp_euler}. For ease of implementation, the spatial domain is fixed to $0<r<r_0$. At each time step $t^n$, the location of detonation front is calculated as $r^* = U_{CJ}t^n$. If the front crosses a new grid point, the state variables at that point are updated to the CJ state values to enforce the boundary conditions. The numerical fluxes are calculated only within the burnt gas region, $0<r<r^*$. This simulation concludes at $t = r_0/U_\text{CJ}$, when the explosive is completely burnt.
	\begin{table}[!htbp]
		\centering
		\begin{tabular}{llllllllll}
			\toprule
			Material   &          EOS         & \multicolumn{8}{l}{Parameters} \\ 
			\\
			\multirow{2}{*}{Burnt gas} & \multirow{2}{*}{JWL} & $\rho_0~(g/mm^3)$ & $p_0~(kPa)$ & $A_1~(GPa)$ & $A_2~(GPa)$ & $R_1$ & $R_2$ & $\omega$ & $q_{\text{det}}~(mm^2/s^2)$  \\ \cline{3-10} 
			&                   &   
			$1.63\times10^{-3}$ & $100$ & $371.2$ & $3.21$ & $4.15$ & $0.95$ & $0.3$ & $4.184\times10^{12}$ 
			\\
			\\
			\multirow{2}{*}{Air} & \multirow{2}{*}{Perfect gas} & $\rho_0~(g/mm^3)$ & $p_0~(kPa)$ & $\gamma$ &  &  &  &  &   \\ \cline{3-5} 
			&                   &   
			$1.177\times10^{-6}$ & $100$ & $1.4$ &  & &  &  &
			\\
			
			\bottomrule
		\end{tabular}
		\caption{Equations of state (EOS) for burnt gas and air and their parameter values \citep{osti_6530310}.}
		\label{tab:eos_parameters}
	\end{table}

	\subsection{Stage 2: Spherical expansion of burnt gas}\label{sec:gas_expansion}
	
	After Stage 1, the burnt gas comes into contact with the ambient air inside the containment structure. The burnt gas has significantly higher pressure, velocity, and density than the air, generating a shock wave that propagates through the air. Until this shock wave reaches the inner wall of the structure, the flow remains spherically symmetric, though consisting of two distinct materials separated by a sharp interface. This spherical two-phase flow is still governed by Equation~\eqref{eq:sp_euler}, except that the spatial domain extends to $(0,~r_1)$, where $r_1$ is the radial coordinate of a point near the structure. Again, $u_\text{r}=0$ is imposed at $r=0$ as a boundary condition. Since the analysis ends just before the shock wave reaches $r_1$, no boundary conditions are needed at $r=r_1$. The burnt gas is initialized using the final solution from Stage 1. The ambient air is modeled using the perfect gas EOS, 
	\begin{equation}
		p=(\gamma-1)\rho e,
		\label{eq:pg}
	\end{equation}
	with $\gamma = 1.4$. In all the test cases presented in this paper, the state variables of air are initialized with $u_0=0$, $\rho_0 = 1.177\times 10^{-6}~\text{g}/\text{mm}^3$ and $p_0 = 100~\text{kPa}$ (Table~\ref{tab:eos_parameters}).
	
	The burnt gas -- air material interface is assumed to be a free surface advected by the fluid flow. Initially, it is located at $r_0$. Its motion is tracked implicitly using the level-set method. Specifically, we introduce a new full-field variable $\phi$, and initialize it to be the signed distance from each point in the domain to the initial location of the material interface. To capture interface motion, we solve the advection equation,
	\begin{equation}
		\dfrac{\partial \phi}{\partial t} + u_\text{r}\dfrac{\partial \phi}{\partial r}=0,~0<x<r_1.
		\label{eq:level_set}
	\end{equation}
	
	At any time $t$, the location of the interface is given by $\Gamma(t)=\{r:~\phi(r,~t)=0\}$. 
	
	The Euler equations are solved using FIVER. Specifically, the numerical fluxes between control volumes that are separated by the material interface are computed by means of solving a 1D fluid-fluid Riemann problem (Figure~\ref{fig:model_problem}). Details of FIVER and the level set method can be found in \cite{main_implicit_2014, wang_computational_2015, main_enhanced_2017, ma_computational_2022}.
	
	
	\vspace{3mm}
	\noindent{\it Remark:} A simpler method is to model the ambient air using the same JWL EOS, which leads to a single-phase flow across the entire spatial domain. This eliminates the need for FIVER  at this stage, and interface tracking becomes optional --- only necessary if the dynamics of the burnt gas ``bubble'' is of interest. This simplification can be justified by the fact that the density of air is much lower than $\rho_0$ in the JWL EOS~\eqref{eq:jwl}, which represents the density of the solid explosive. As a result, when JWL is used to model the ambient air, the exponential terms in the EOS are negligible. Essentially, the EOS degenerates to perfect gas, with Gr\"uneisen parameter $\omega = \gamma - 1$. The error introduced in this simplification arises from setting $\omega = 0.3$ based on TNT, while the correct value for air is $\gamma - 1 = 0.4$. In Section~\ref{sec:fsi_results}, we present a test case that demonstrates the minimal impact of this approximation on the flow results.

	\subsection{Stage 3: Fluid-structure interaction}\label{sec:blast_mitigation}
	
\begin{figure}[!htb]
		\centering\includegraphics[width=150mm]{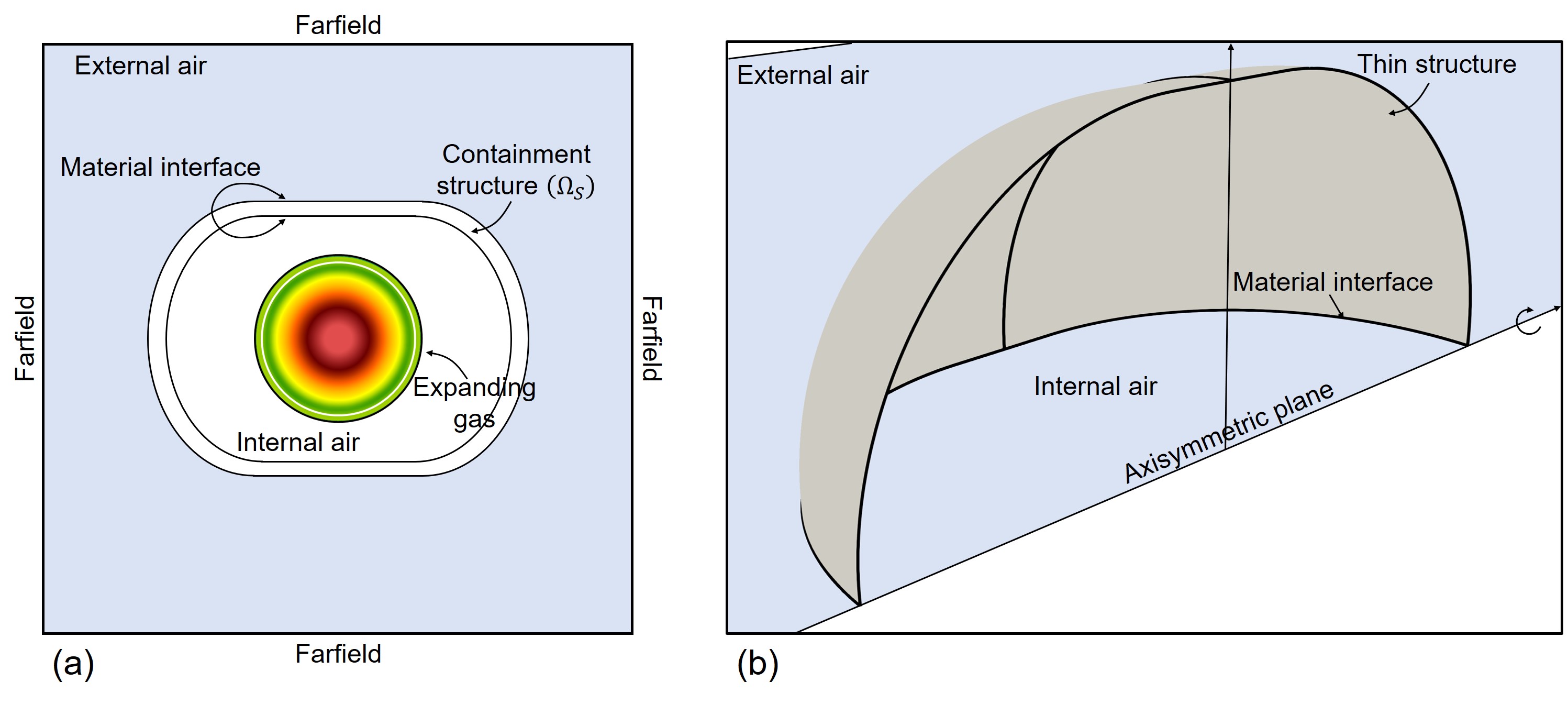}
		\caption{Schematic of the coupled fluid-structure interaction analysis in Stage 3. (a) The physical domain represented in the axisymmetric plane. (b) The 2D-axisymmetric fluid domain and the 3D solid structure.}
		\label{fig:flattend_model_problem}
	\end{figure}

	Figure~\ref{fig:flattend_model_problem} illustrates  the simulation setup for this stage. We assume the containment structure has cylindrical symmetry, but not spherical symmetry. Specifically, its cross-section is circular, while the planform has a general closed shape, as depicted in Figure~\ref{fig:flattend_model_problem}(a). This assumption is valid for most pressure vessels in real-world applications. The fluid domain includes both the gas inside the chamber, and a volume of air outside. Because the volume of fluid is much larger than that of the solid, the total computational cost is dominated by the fluid dynamics solver. To reduce this cost, we formulate the fluid governing equations in 2D using cylindrical symmetry, while maintaining a 3D model for the structural analysis. 
	
	Within this stage, we model both the burnt gas and the air inside the containment structure using the JWL EOS~\eqref{eq:jwl}, and initialize them with the density, velocity, and pressure results obtained from Stage 2. As mentioned in Section~\ref{sec:gas_expansion}, modeling air using the JWL EOS can be justified by its low density compared to the solid explosive. Moreover, the energy carried by the ambient air is much smaller than that of the burnt gas. Therefore, the loss of total energy due to the change of EOS is negligible. Outside the containment structure, the air is modeled using the perfect gas EOS~\eqref{eq:pg}.
	
	
	
	\subsubsection{Fluid governing equations and numerical discretization}\label{sec:fvm}
	
	Let $\Omega_\text{F}(t)$ represent the fluid domain at time $t$. Because of transient structural deformation, the boundary of $\Omega_\text{F}(t)$ is unknown {\it a priori} and must be determined by solving the fluid-structure coupled problem. Moreover, given the axisymmetric nature of the containment structure studied in this work, the fluid flow is assumed to be axisymmetric, with no circumferential velocity $\big(u_\theta\big)$. Under this condition, the Euler equations for compressible inviscid flows are given by
	
	\begin{equation}
		\dfrac{\partial\textbf{W}}{\partial t}+\dfrac{\partial\mathcal{F}(\textbf{W})}{\partial r}+\dfrac{\partial\mathcal{G}(\textbf{W})}{\partial z}=-\dfrac{1}{r}\mathcal{S}(\textbf{W})
		\quad \textbf{x}\in\Omega_\text{F},
		\label{eq:axisym_euler}
	\end{equation}
	where $\displaystyle \textbf{W}=\big[\rho,~\rho u_\text{r},~\rho w,~E\big]^T$ denotes the conservative state variables, with $u_\text{r}$ and $w$ representing the radial and axial velocity components. The flux functions $\mathcal{F}$ and $\mathcal{G}$ and the right-hand side source term $\mathcal{S}$ are given by
	\begin{equation}
		\mathcal{F}\big(\textbf{W}\big)=
		\begin{bmatrix}
			\rho u_\text{r}\\
			\rho u_\text{r}^2+p\\
			\rho u_\text{r}w\\
			u_\text{r}(E +p)
		\end{bmatrix},\quad
		\mathcal{G}\big(\textbf{W}\big)=
		\begin{bmatrix}
			\rho w\\
			\rho u_\text{r}w\\
			\rho w^2+p\\
			w(E +p)
		\end{bmatrix},
		\quad
		\mathcal{S}\big(\textbf{W}\big)=
		\begin{bmatrix}
			\rho u_\text{r}\\
			\rho u_\text{r}^2\\
			\rho u_\text{r}w\\
			u_\text{r}(E+p)
		\end{bmatrix}.
		\label{eq:axisym_euler_terms}
	\end{equation}
	
	Let $\Gamma_\text{F}$ denote the fluid domain boundary that is in contact with the solid structure, that is, the fluid-structure interface. Across $\Gamma_\text{F}$, normal velocity is assumed to be continuous, which leads to a Dirichlet boundary condition for the fluid,
	\begin{equation}
		\vec{v}\cdot\vec{n}_\Gamma=\Dot{\vec{u}}\cdot\vec{n}_{\Gamma},
		\label{eq:kinematic_bc}
	\end{equation}
	where $\displaystyle \vec{v}=\big[u_\text{r},~w\big]^T$ represents the fluid velocity. $\Dot{\vec{u}}$ is the structural velocity, and $\vec{n}_{\Gamma}$ denotes the direction normal to $\Gamma_\text{F}$. 
	
	Using the embedded boundary method, we discretize Equation~\eqref{eq:axisym_euler} on a fixed finite volume mesh that encompasses the fluid domain and the solid structure. The mesh is not expected to have a set of nodes that lie on the fluid-structure interface, and hence can be as simple as a uniform Cartesion grid. Within each control volume $C_i$,
	\begin{equation}
		\dfrac{d\textbf{W}_i}{dt}=-\dfrac{1}{|C_i|}\sum_{C_j\in N(i)}\Big(
		\int_{\partial C_{ij}}\mathcal{F}\big(\textbf{W}\big) n_{ij}^rdS+\int_{\partial C_{ij}}\mathcal{G}\big(\textbf{W}\big) n_{ij}^zdS
		\Big)-\dfrac{1}{r_i}\mathcal{S}\big(\textbf{W}_i\big),
		\label{eq:discretized_euler}
	\end{equation}
	where $\textbf{W}_{i}$ is the average values of the conservative state variables in the control volume. $N(i)$ is the set of neighboring control volumes. For each control volume $C_j\in N(i)$, $\partial C_{ij}$ represents the shared boundary between $C_i$ and $C_j$. The outward normal at $\partial C_{ij}$ has radial and axial components $n^r_{ij}$ and $n^z_{ij}$, respectively. $\norm{C_i}$ denotes the volume of $C_i$. 
	
	
	We use the FIVER method to restrict flow solution within the actual fluid domain $\big(\Omega_\text{F}\setminus\Omega_{\text{S}}\big)$ 
	and to enforce the boundary condition~\eqref{eq:kinematic_bc}. Specifically, for each pair of neighboring control volumes, $C_i$ and $C_j$:
	\begin{itemize}
		\item[(i)] If $C_i$ and $C_j$ both belong to the fluid domain $\Omega_\text{F}$ and are not separated by $\Gamma_\text{F}$, the standard finite volume method is utilized to compute the mass, momentum, and energy fluxes across $\partial C_{ij}$.
		\item[(ii)] If $C_i$ and $C_j$ are both covered by the solid structure, no fluxes are computed and the state variables within both cells remain unchanged.
		\item[(iii)] If $C_i$ and $C_j$ are separated by $\Gamma_\text{F}$, and one of them is covered by the solid structure, a 1D fluid-structure Riemann problem is constructed locally (Figure~\ref{fig:model_problem}). This problem features a constant wall velocity specified by the structural dynamics solver. The solution of this Riemann problem is used to compute the fluxes across $\partial C_{ij}$, updating only the state variables in the control volume that belongs to the fluid domain.
		\item[(iv)] If $C_i$ and $C_j$ are separated by the interface and neither is covered by the solid structure, two separate 1D fluid-structure Riemann problems are constructed for $C_i$ and $C_j$. Their solutions are used to evaluate the flux functions, which are then employed to update the state variables in $C_i$ and $C_j$, respectively.
	\end{itemize}
	

	When $\Gamma_\text{F}$ sweeps through a control volume $C_i$, the Equation of State (EOS) is updated to reflect the appropriate fluid material. FIVER then employs the interfacial fluid state, determined by solving a local one-dimensional fluid-structure Riemann problem, to correct the conservative state variables within the affected control volume. D	etails of FIVER for fluid-structure interaction problems can be found in~\cite{farhat_dynamic_2013, main_enhanced_2017, cao_shock_2021, ma_computational_2022}.

	\subsubsection{Finite-strain structural dynamics}\label{sec:fea}
	
	Unlike the fluid, the structural governing equations are formulated using the Lagrangian reference frame. Let $\Omega_\text{S}(t)$ be the configuration of the explosion containment structure at time $t$. The dynamic equilibrium equations are given by
	\begin{equation}
		\rho_\text{S}\frac{\partial^2\vec{u}}{\partial t^2}=\frac{\partial}{\partial \vec{X}}\left(
		\vec{S}+\vec{S}\cdot\frac{\partial\vec{u}}{\partial\vec{X}}^T
		\right), \qquad \forall~\vec{X}\in\Omega_\text{S}(0),
		\label{eq:solid_dynam}
	\end{equation}
	assuming there are no body forces. Here, $\rho_\text{S}$, $\vec{X}$, and $\vec{S}$ denote the structural density, material coordinates, and the second Piola-Kirchhoff stress tensor, respectively. $\displaystyle\vec{u}=\big[u_\text{X},~u_\text{Y},~u_\text{Z}\big]^T$ is the displacement vector. We use the second Piola-Kirchhoff stress and its conjugate Green-Lagrange strain $\big(\vec{E}\big)$ as the stress and strain measures because of their frame independence. $\vec{S}$ and $\vec{E}$ are related through the material's constitutive model,
	\begin{equation}
		\vec{S}=\frac{\partial W(\vec{X}, t)}{\partial \vec{E}},
		\label{eq:matequations}
	\end{equation}
	where $W$ is the strain energy potential. For the elasto-plastic metallic materials used in this work, the stress-strain relation is characterized by a bi-linear model. In this case, Equation ~\eqref{eq:matequations} reduces to
	\begin{equation}
		\vec{S}=\vec{C}:\left(\vec{E} - \vec{E}_\text{P}\right),
	\end{equation}
	where $\vec{C}$ denotes the fourth-order constitutive tensor, and $\vec{E}_\text{P}$ is the plastic part of the strain measure, computed using the radial return algorithm~\cite{simo_computational_2006}.
	
	
	Let $\Gamma_\text{S}$ denote the boundary of the structure that is in contact with the fluid. We enforce force equilibrium,
	\begin{equation}
		\boldsymbol{\sigma}_\text{S}\cdot\vec{n}_{\Gamma} = p\vec{n}_{\Gamma},
		\label{eq:traction_bc}
	\end{equation}
	where $\vec{n}_\Gamma$ is the direction normal to the boundary $\Gamma_\text{S}$ and $\boldsymbol{\sigma}_\text{S}$ denotes the Cauchy stress tensor of the structure. It is related to the second Piola-Kirchhoff stress, $\vec{S}$, by $\displaystyle \boldsymbol{\sigma}_\text{S}=J^{-1}\vec{F}\cdot\vec{S}\cdot\vec{F}^T$. Here, $\vec{F}$ is the deformation tensor and $J=\text{det}\big(\vec{F}\big)$. 
	
	In this work, Equation~\eqref{eq:solid_dynam} is projected to a discrete space using the standard Galerkin finite element method, with the displacement field approximated by
	\begin{equation}
		\vec{u}\left(\vec{X}, t\right)=\sum_{I=1}^n N_I\left(\vec{X}\right)\vec{u}_I(t),
		\label{eq:shape_func}
	\end{equation}
	where $n$ represents the number of nodes, $N_I\left(\vec{X}\right)$ are linear shape functions, and $\vec{u}_I$ denotes the nodal displacement vector. Using the trial functions $\delta\vec{u}=N_I\delta \vec{u}_I$ we obtain the finite element equations:
	\begin{equation}
		\vec{M}\Ddot{\vec{u}}+\vec{F}^{int}=\vec{F}^{ext},
		\label{eq:finite_dynam}
	\end{equation}
	where $\vec{M}$ is the mass matrix, and $\vec{F}^{int}$ denotes the interface forces due to material constitutive behavior. $\vec{F}^{ext}$ represents the external loads, in this case given by the fluid through the equilibrium interface condition:
	\begin{equation}
		\vec{F}^{ext}_I=\int_{\Gamma_\text{S}}N_I p\vec{n}_{\Gamma}  dS.
		\label{eq:external_vector}
	\end{equation}
	
	%
	%

	\subsubsection{Numerical method for fluid-structure coupling}\label{sec:time_marching}
	
	The fluid and structural governing equations are coupled through the two interface conditions, \eqref{eq:kinematic_bc} and \eqref{eq:traction_bc}, and the shared boundary,
	\begin{equation}
		\Gamma_\text{F}(t) = \overline{\Omega_\text{F}(t)} \cap \overline{\Omega_{\text{S}}(t)} = \Gamma_{\text{S}}(t),\quad \forall t>0,
		\label{eq:shared_interface}
	\end{equation}
	where the overline indicates set closure.
	
	In this work, we adopt a partitioned procedure to couple the fluid and structural dynamics solvers. The time grids of the two solvers are offset by half a step to achieve second-order accuracy~\cite{farhat_robust_2010}. In an arbitrary time step $n$, the structural solver sends the current location and velocity of $\Gamma_\text{S}$ to the fluid solver. Then, the fluid solver advances one time step, with
	\begin{align}
		\Gamma_\text{F}(t) &= \Gamma_\text{S}(t^{n}_\text{S}),\\
		\vec{v}(t)\cdot\vec{n}_\Gamma &= \dot{\vec{u}}(t^{n}_\text{S})\cdot\vec{n}_\Gamma,\quad t^{n}_\text{F}\leq t < t^{n+1}_\text{F},
	\end{align}
	where $t^{n}_\text{S}$ and $t^{n}_\text{F}$ represent time points in the structural and fluid solvers, respectively.
	
	$\Gamma_\text{S}(t^{n}_\text{S})$ is treated as an embedded boundary in the fluid solver, represented by a surface mesh. It is tracked within the non-interface conforming fluid mesh using a collision-based algorithm, as described in~\cite{wang_computational_2012}. The intersections between the edges in the fluid mesh and the surface elements are identified. The status of each cell in the fluid mesh --- that is, which material subdomain it belongs to --- is also determined. These information are utilized to compute numerical fluxes, as described in Section~\ref{sec:fvm}.
	
	After completing the time step, the fluid solver computes the distributed pressure loads on the embedded surface mesh, and sends them to the structural solver. The structural solver advances one time step with
	\begin{equation}
		\boldsymbol{\sigma}_\text{S}(t)\cdot\vec{n}_{\Gamma} = p(t^{n+1}_\text{F})\vec{n}_{\Gamma},\quad t^{n}_\text{S}\leq t < t^{n+1}_\text{S}.
	\end{equation}
	
	In this work, the semi-discretized fluid governing equations are integrated using a third-order explicit Runge-Kutta (RK) method, while the structural governing equations are integrated with a second-order central difference method. Given that the fluid governing equations are solved on a 2D mesh and the structural model is 3D, the solver communications are carried out following cylindrical mappings.

	\section{Verification tests}\label{sec:verification}
	
	We employ the finite volume method implemented in Aero-F \cite{aerof} to simulate the detonation process (Stage 1). The numerical methods for simulating the fluid and structural dynamics in Stages 2 and 3 are implemented in M2C \cite{m2c} and Aero-S \cite{aeros}, respectively. All three solvers are open-source, developed in C++, and parallelized using the Message Passing Interface (MPI). The simulations presented in this paper were conducted on the Tinkercliffs High-Performance Computing (HPC) cluster at Virginia Tech.

	\subsection{The detonation process}\label{sec:detonation_verification}
	
\begin{figure}[!htb]
		\centering\includegraphics[width=90mm]{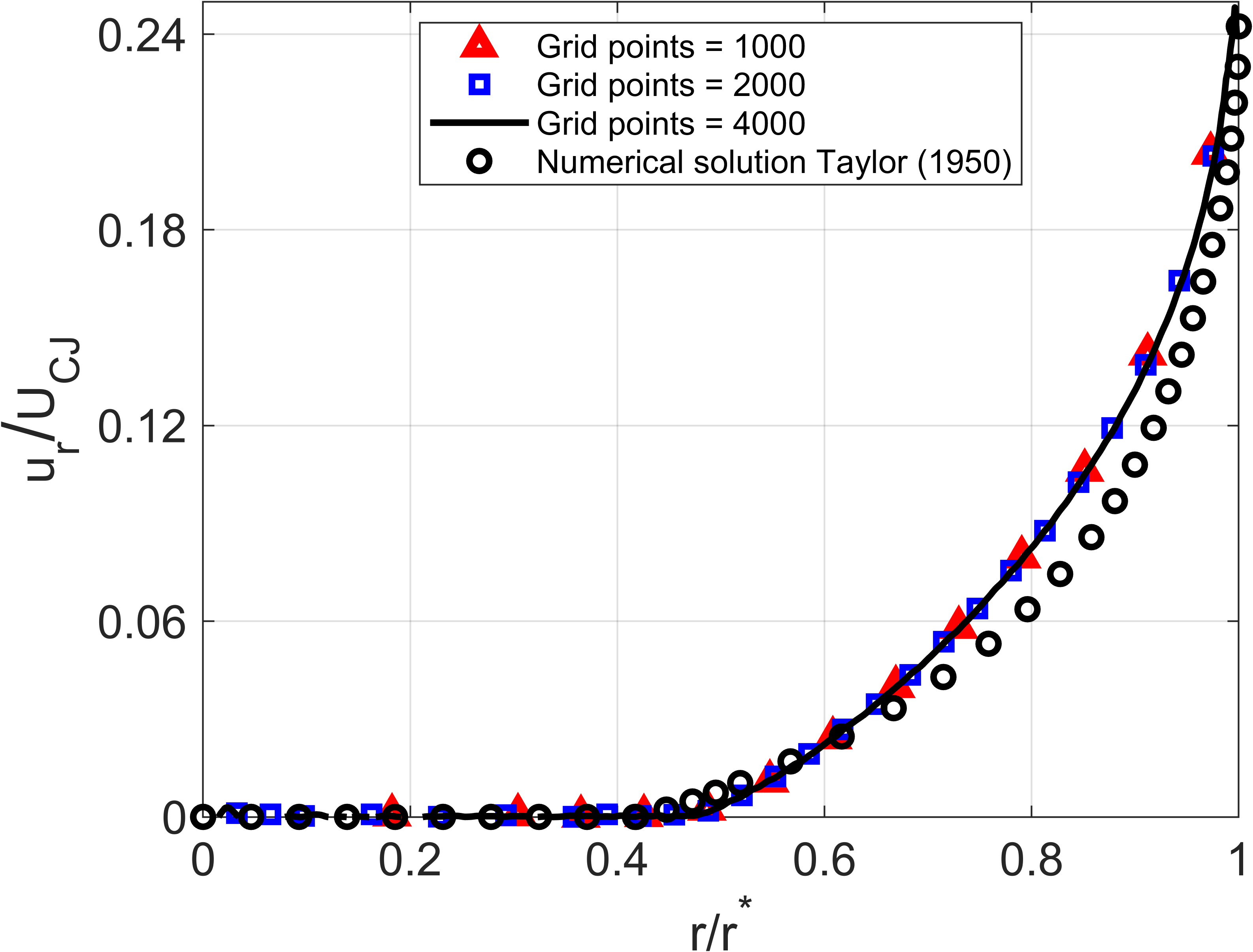}
		\caption{The detonation process: Velocity distribution behind the spherical detonation wave, scaled using the detonation wave speed ($U_{\text{CJ}}$) and the front's radial distance ($r^*$).}
		\label{fig:burning_convergence}
	\end{figure}

	We verify the spherical detonation analysis presented in Section~\ref{sec:detonation} using an example problem discussed by Taylor in \cite{taylor_dynamics_1950}, which employs $U_{\text{CJ}} = 6.38\times10^6~\text{mm}/\text{s}$ and $\rho_{CJ}=2\times10^{-3}~\text{g}/\text{mm}^3$. However, we set $\rho_0=1.51\times10^{-3}~\text{g}/\text{mm}^3$ according to Taylor~\cite{taylor_dynamics_1950}, and compute the CJ states using the Rankine-Hugoniot jump conditions and the JWL EOS, with $q_{\text{det}}=4.184~\text{mm}^2/\text{s}^2$ (TNT). The computed CJ values are $\rho_{\text{CJ}} = 2.0583\times10^{-3}~\text{g}/\text{mm}^3$, $U_{\text{CJ}} = 7.1473\times10^6~\text{mm}/\text{s}$, and $p_{\text{CJ}} = 2.0549\times10^{10}~\text{Pa}$.
	Figure~\ref{fig:burning_convergence} presents the numerical solutions obtained with different mesh resolutions, in comparison with the reference result. Mesh convergence is clearly evident. The nondimensionalized velocity, $u_r/U_{CJ}$, at detonation front closely matches the reference result. The rarefaction wave profile is similar to the reference, but not exactly the same. The discrepancy is expected since Taylor used the perfect gas EOS to model the burnt gas~\cite{taylor_dynamics_1950}.
	
	\subsection{Spherical gas expansion}\label{sec:gas_expansion_verification}
	
	To verify the spherical fluid simulation in Stage 2, we conduct four tests with the mass of explosive charge, $m_0 = 226.8~\text{g}$, $453.6~\text{g}$, $680.4~\text{g}$, and $907.2~\text{g}$, respectively. In all the tests, the explosive is assumed to be TNT and modeled using the JWL EOS, with parameter values specified in Table~\ref{tab:eos_parameters}. Because the detonation process (Stage 1) yields self-similar solutions, it only needs to be simulated once to generate initial conditions for all the four tests.
	
\begin{figure}[!htb]
		\centering\includegraphics[width=150mm]{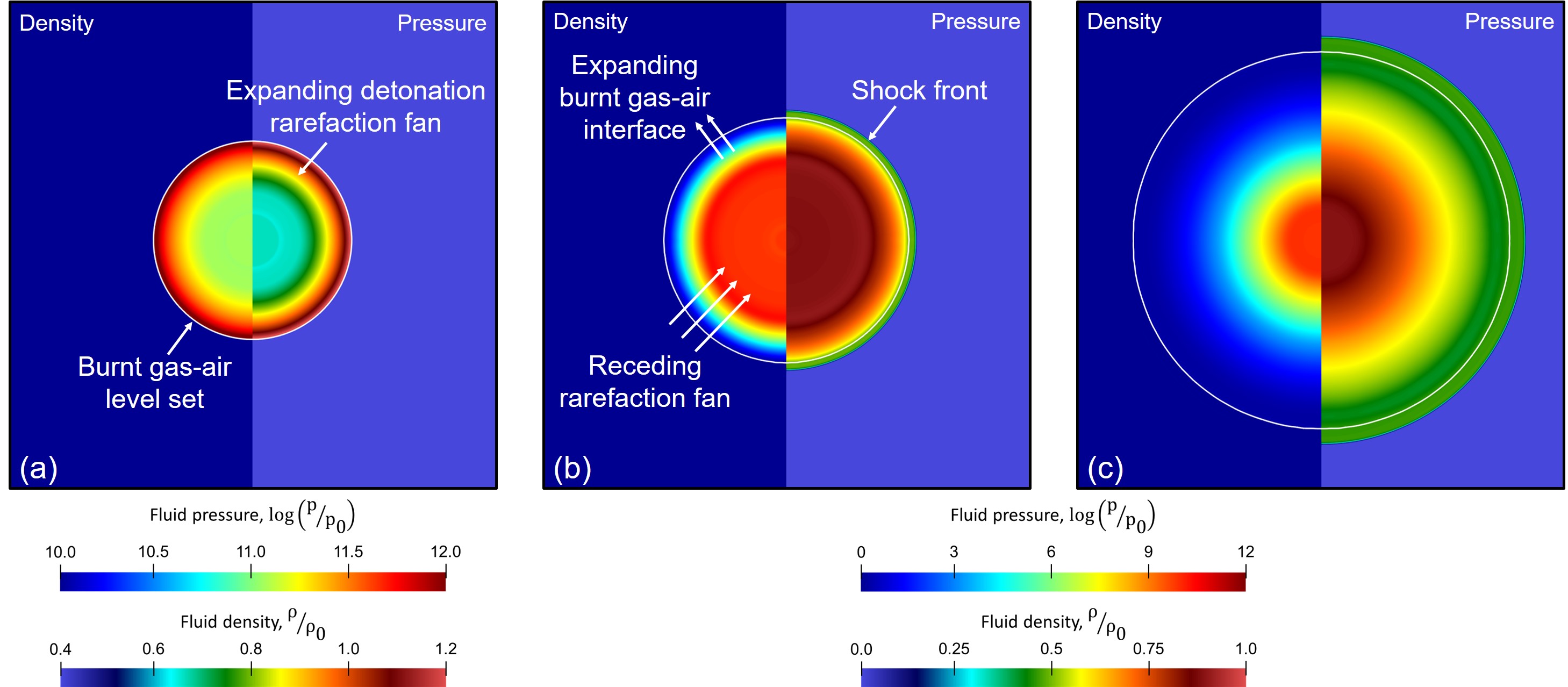}
		\caption{Spherical expansion of burnt gas: simulation results ($m_0 = 226.8~\text{g}$) at (a) $t=0~s$, (b) $t=1.5\times10^{-6}~s$, and (c) $t=6.0\times10^{-6}~s$.}
		\label{fig:density_snapshots}
	\end{figure}

	Figure~\ref{fig:density_snapshots} presents three density and pressure snapshots obtained with $m_0 = 226.8~\text{g}$. The initial discontinuities across the burnt gas - air interface generate a shock wave that propagates into the ambient air. They also create a backward moving rarefaction fan, which flattens the expanding rarefaction fan formed in the detonation process (Figure~\ref{fig:burning_convergence}). The material interface itself expands outwards, as the burnt gas is moving with high kinetic energy into a medium that is stationary. Given that the shock wave is relatively weak, most of the detonation energy is contained within the spherical ``bubble'' of burnt gas.
	
\begin{figure}[!htb]
		\centering\includegraphics[width=75mm]{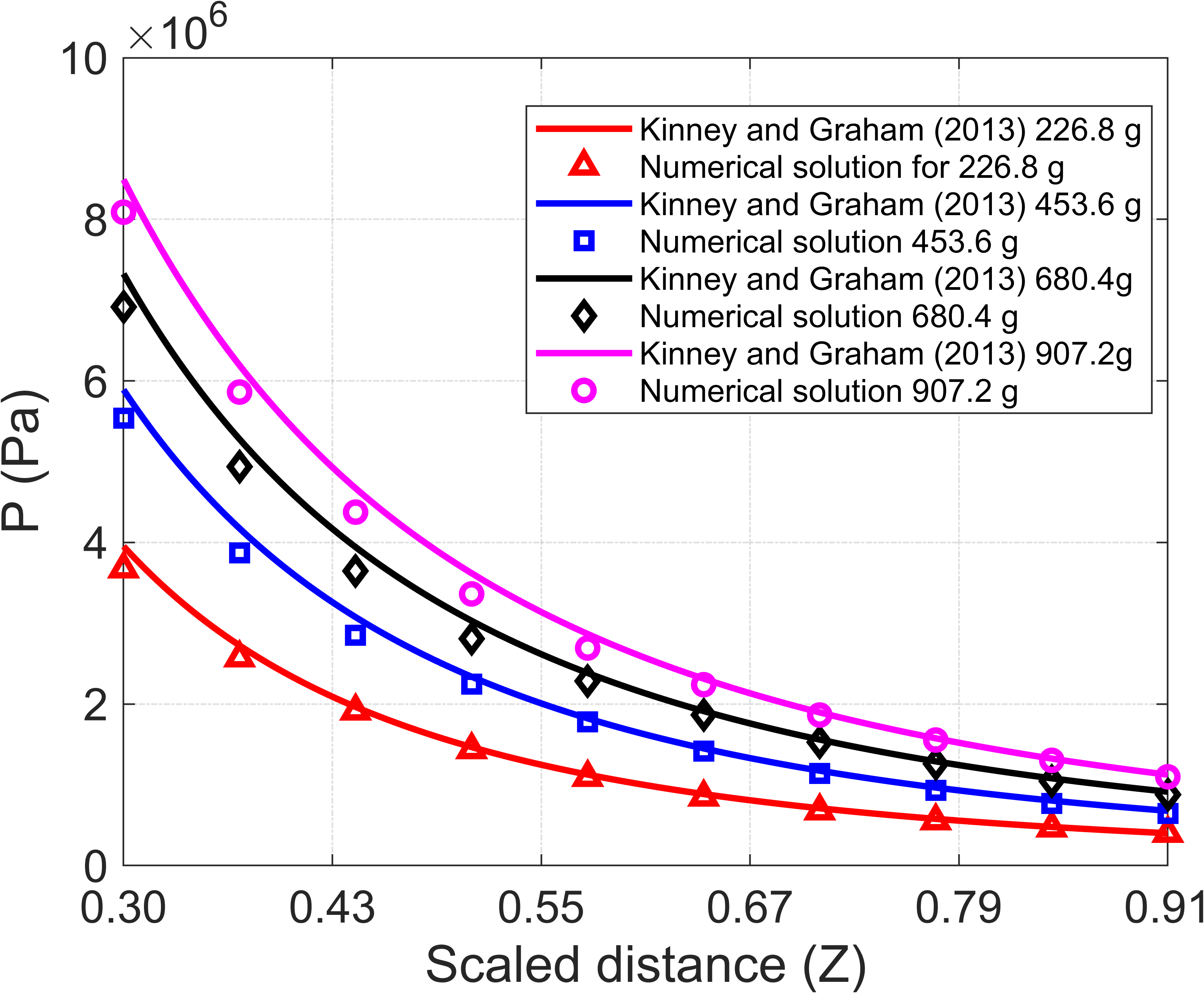}
		\caption{Spherical expansion of burnt gas: Comparison of maximum over-pressures obtained from numerical simulations and an empirical model. }
		\label{fig:overpressure_comparison}
	\end{figure}

	Ten pressure probes are uniformly distributed within the spatial domain. Figure~\ref{fig:overpressure_comparison} presents the maximum pressure values at these locations, captured when the ``bubble'' of burnt gas crosses the probes. These values are compared with the equation in~\cite{kinney_explosive_2013} for over-pressures from explosions, 
	\begin{equation}
		p \big(Z\big)=\dfrac{808\times\Big(1+\Big(\dfrac{\text{Z}}{4.5}\Big)^2\Big)}{\sqrt{
				\Big(1+\Big(\dfrac{\text{Z}}{0.048}\Big)^2\Big)\times
				\Big(1+\Big(\dfrac{\text{Z}}{0.32}\Big)^2\Big)\times
				\Big(1+\Big(\dfrac{\text{Z}}{1.35}\Big)^2\Big)
		}}\times10^5~\text{Pa},
		\label{eq:kinney_pressure}
	\end{equation}
	where $Z$ represents a scaled distance $\big($dimension: [length]$/$[mass]$^{1/3}$$\big)$.
	
	Figure~\ref{fig:overpressure_comparison} shows that for all four test cases, the numerical results are in close agreement with values obtained using the empirical model.
	
	\subsection{Blast loaded plates}\label{sec:blast_plates}
	
	\begin{figure}[!htb]
		\centering\includegraphics[width=120mm]{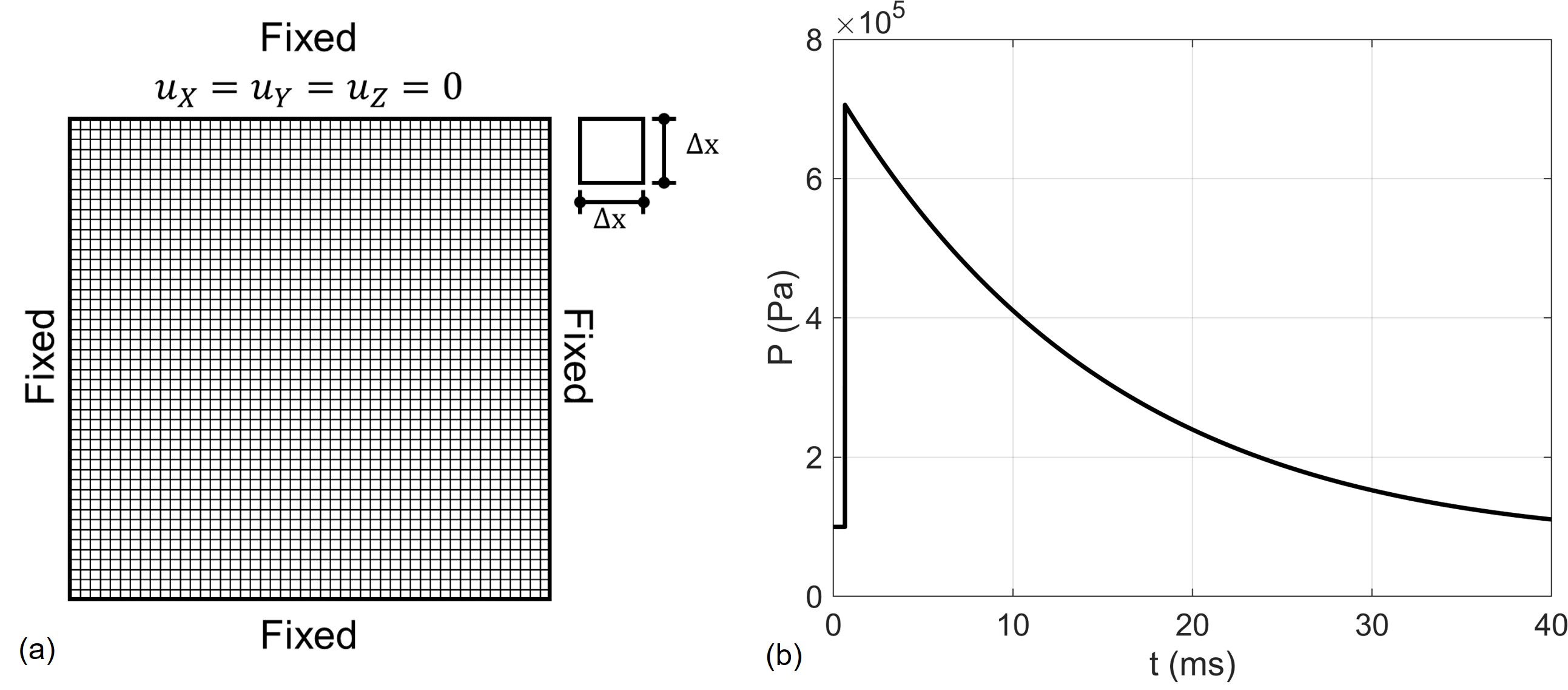}
		\caption{Blast loaded plate. (a) Solid mesh with boundary conditions. (b) Pressure pulse obtained from Equation~\eqref{eq:friedlander}.}
		\label{fig:plate_problem}
	\end{figure}
	
	A numerical study of square plates subjected to blast loading is performed to assess the accuracy of the structural dynamics solver. The problem is a simplified representation of the shock-tube experiments conducted by \cite{aune_shock_2016}. Specifically, we replicate their Test D77-15, in which a deformable Docol 600 DL plate is exposed to a shock load generated by a 15 bar firing pressure. The plate is discretized using the shell finite elements \cite{belytschko_explicit_1984}. The exposed area is of size $\text{300 mm}\times\text{300 mm}$ with a plate thickness of $0.8$ mm. The pressure load is generated using equation
	\begin{equation}
		p_r(t)=p_a+p_{r,max}\left(1-\frac{t-t_a}{t_{d+}}\right)\text{exp}\left(\frac{-b(t-t_a)}{t_{d+}}\right), \quad t_a<t<t_a+t_{d+},
		\label{eq:friedlander}
	\end{equation}
	
	\begin{figure}[!htb]
		\centering\includegraphics[width=70mm]{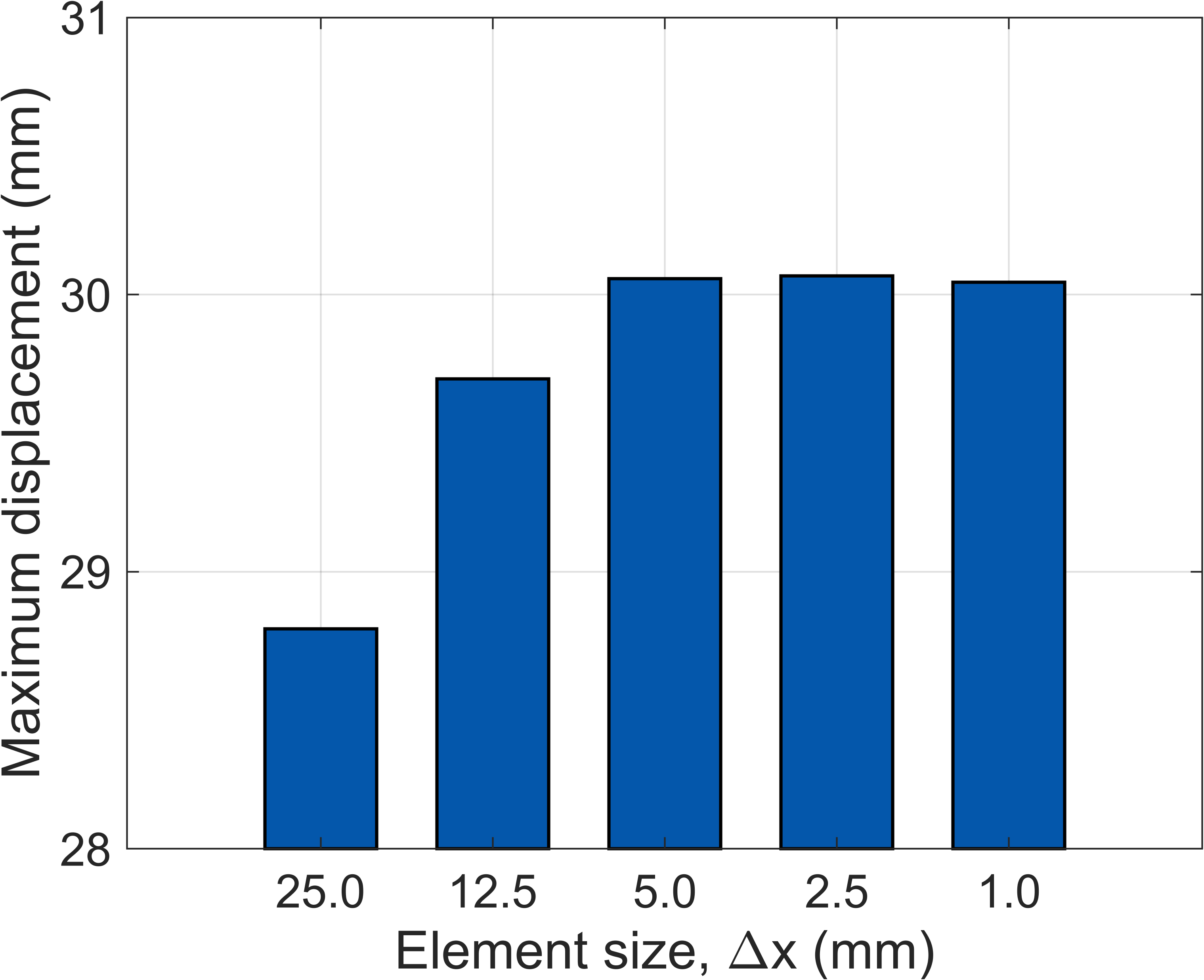}
		\caption{Blast loaded plate: mesh sensitivity analysis.}
		\label{fig:shell_convergence}
	\end{figure}
	
	which was presented by Friedlander in \cite{friedlander_diffraction_1946}. We set $p_a=99.3~\text{kPa}$, $p_{r,max}=606.6~\text{kPa}$, $t_a=0.64\times10^{-3}~s$, $t_{d+}=44.1~\text{s}$, and $b=2.025$, according to \cite{aune_shock_2016}. The numerical setup along with the pressure time history are shown in Figure~\ref{fig:plate_problem}.
	
\begin{figure}[!htb]
		\centering\includegraphics[width=70mm]{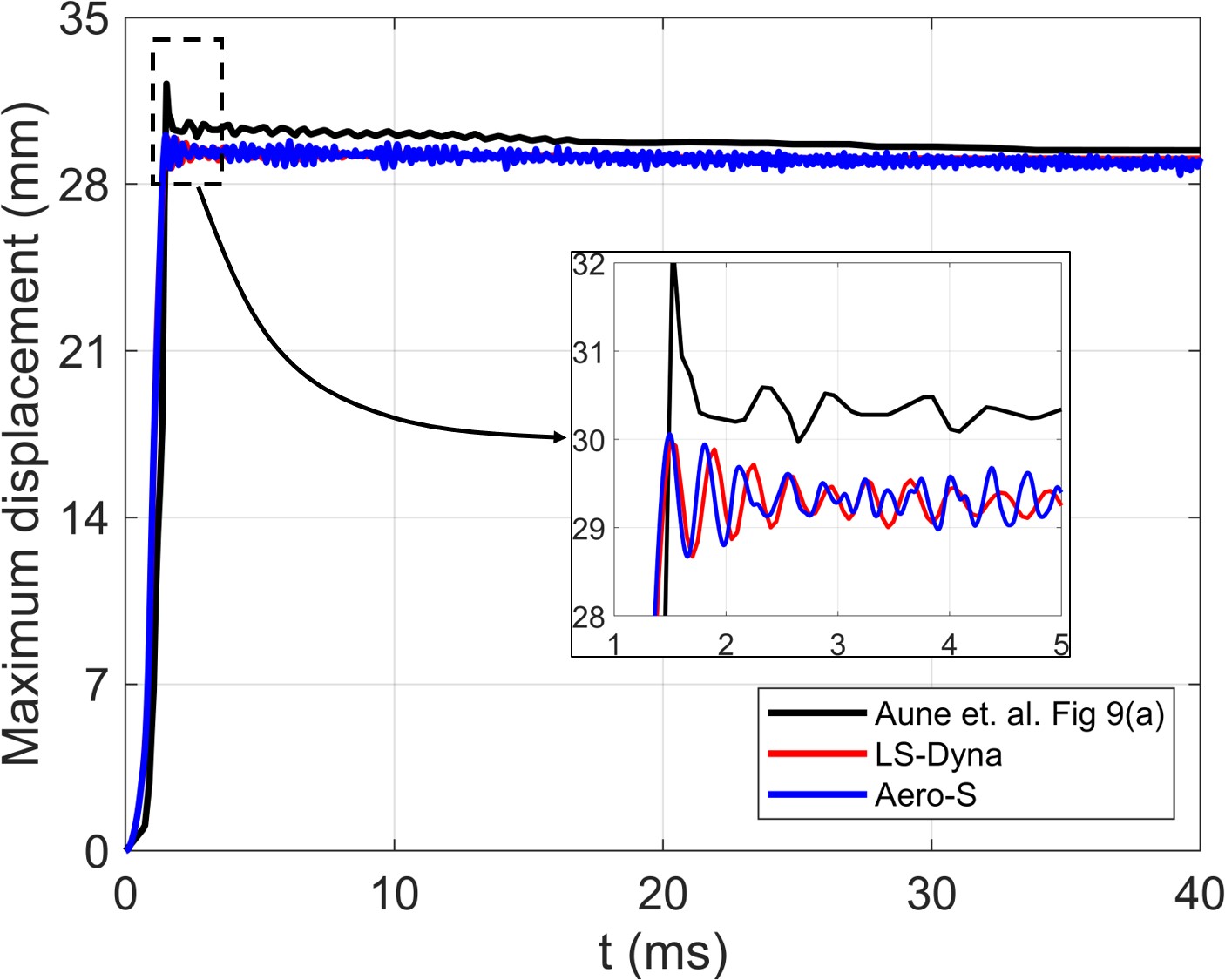}
		\caption{Blast loaded plate: time-history of mid-point displacement magnitude.}
		\label{fig:shell_verify}
	\end{figure}

	A mesh convergence analysis is first conducted, with element size $\Delta x$ varied between $25~\text{mm}$ and $1~\text{mm}$. Figure~\ref{fig:shell_convergence} presents the maximum plate deflection obtained from all the test cases. It can be observed that as the element size decreases, the solution converges towards $30.04~\text{mm}$. This converged numerical solution differs from the experimental data by approximately $5.6\%$ (Figure~\ref{fig:shell_verify}). The discrepancy, at least part of it, can be attributed to the way plate boundaries are treated. Our numerical study assumes a fixed edge whereas in the experiment, the plate was held in place by clamping two thick rigid plates together, allowing it to slip and tear near clamping bolts.
	
\begin{figure}[!htb]
		\centering\includegraphics[width=125mm]{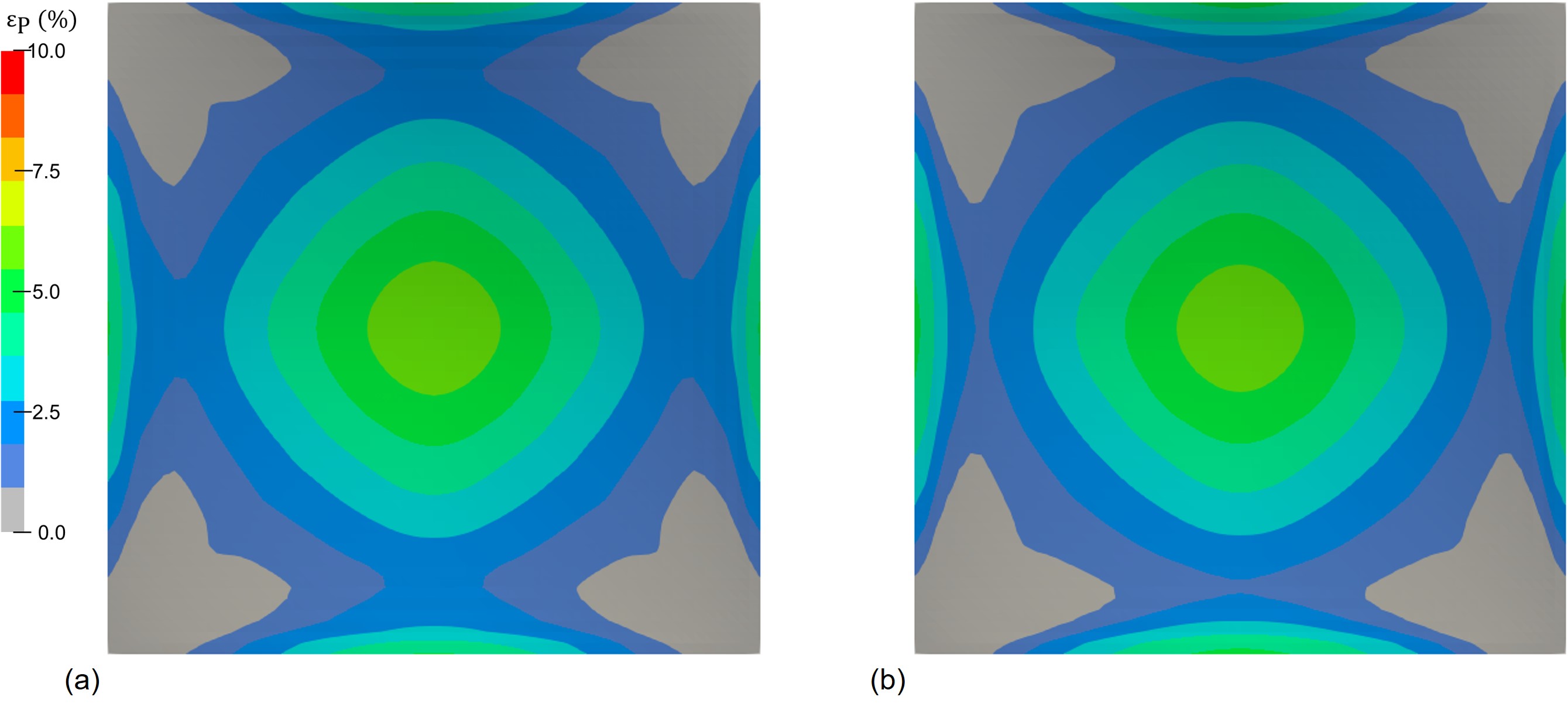}
		\caption{Blast loaded plate: Predicted effective plastic strain ($\varepsilon_\text{P}$) at $t = 40~\text{ms}$. (a) LS-Dyna. (b) Aero-S.}
		\label{fig:dyna_vs_aero}
	\end{figure}

	Additionally, we have also simulated the same problem using LS-Dyna. The pressure time history~\eqref{eq:friedlander} is provided to the solver through a *DEFINE\_CURVE card. Figures~\ref{fig:shell_verify} and~\ref{fig:dyna_vs_aero} confirm that the result obtained from our structural dynamics solver (Aero-S) is similar to LS-Dyna's output. 
	
	In external explosion events, such as the one illustrated here, the initial impact is of utmost significance. Depending on the incident shock pulse, the plate may respond elastically or elasto-plastically. In this case, the shock is strong enough to drive the plate material into the plastic regime, as shown in Figure~\ref{fig:dyna_vs_aero}. Following the incident shock pulse, the decaying nature of the pressure wave causes the structure to unload and reload elastically, resulting in vibrations around a certain equilibrium configuration, as depicted in Figure~\ref{fig:shell_verify}.
	
	\section{Case study: Lightweight explosion-containment chamber}\label{sec:results}
	
	Previous studies on structural responses to internal explosions have often relied on predefined pressure loads, neglecting the dynamic interaction between structural deformation and the internal fluid flow. We hypothesize that this approach may be inadequate for lightweight, single-use structures that are subject to significant plastic deformations. To test this hypothesis and demonstrate the proposed simulation method, we design a numerical experiment comparing three types of simulations.
	
	\begin{itemize}
		\item \textbf{FSI}: Fluid-structure coupled analysis based on the computational model presented in Section~\ref{sec:blast_mitigation} (Stage 3), with the solution obtained from Stage 2 as the initial condition.
		\item \textbf{FSD}: Decoupled fluid and structural analysis. The fluid flow is first simulated within a rigid containment structure that serves as a fixed wall boundary. The solution obtained from Stage 2 is used as its initial condition. The time-history of the pressure field on the wall is recorded and subsequently applied as a predefined external load to simulate the dynamics of the containment structure.
		\item \textbf{SD}: Decoupled structural analysis, with a transient pressure load defined using the Friedlander equation~\eqref{eq:friedlander}.
	\end{itemize}
	
	

	\subsection{Model setup}\label{sec:model_setup}
	
\begin{figure}[!htb]
		\centering\includegraphics[width=150mm]{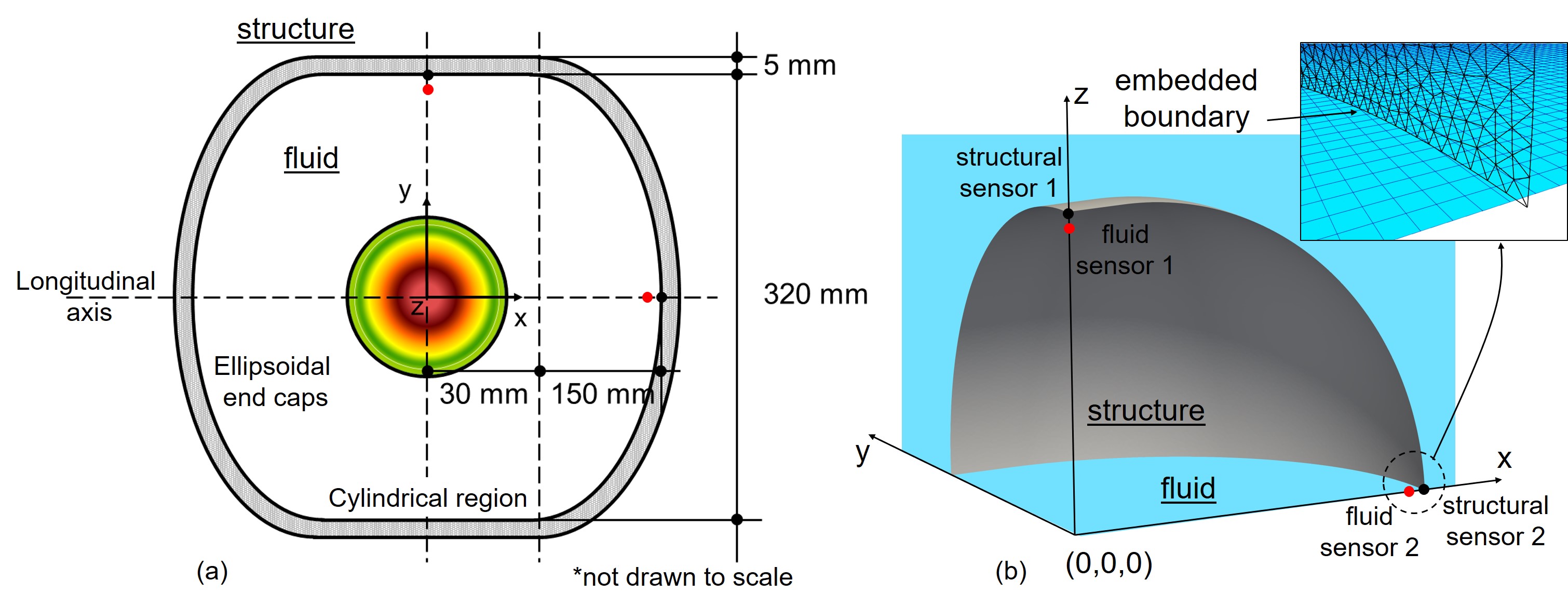}
		\caption{Setup of model problem. (a) Planform view and dimensions of the blast containment structure. (b) Fluid and structural meshes and sensor locations.}
		\label{fig:numerical_setup}
	\end{figure}

	Figure~\ref{fig:numerical_setup} depicts the explosion containment structure used in all three simulations. It is made of steel with the following material properties: density $\rho_\text{S}=7.9\times10^{-3}~\text{g}/\text{mm}^3$, Young's modulus $E=210~\text{GPa}$, Poisson's ratio $\nu=0.3$, and yield strength $\sigma_\text{Y}=355~\text{MPa}$. The material is assumed to be perfectly plastic beyond its elastic limit. We employ the well-known $J_2$ radial return algorithm to model the material's elasto-plastic constitutive behavior.
	
	The geometry of the containment structure follows the standard pressure vessel design, comprising ellipsoidal end caps that are connected by a cylindrical region (Figure~\ref{fig:numerical_setup}(a)). We set the cylindrical region to be $60~\text{mm}$ long, while the ellipsoidal ends have a major and minor radii of $160~\text{mm}$ and $150~\text{mm}$, respectively. The structure is assumed to have no gaps or holes. Since the primary focus of this work is on lightweight single-use explosion containment chambers, the structure is modeled as a thin shell with a thickness of $5~\text{mm}$. In Section~\ref{sec:blast_plates}, we have shown that the shell element implemented in our structural dynamics solver provides good agreement with both LS-Dyna and experimental data for a blast application. Here, we employ the same type of element to discretize the containment structure. Leveraging cylindrical symmetry, only $1/8$ of the structure is simulated. Additional geometric parameters and boundary conditions are shown in Figure~\ref{fig:numerical_setup}.

	
	Two sensors are placed on the structural boundary, where displacement, velocity, and effective plastic strain are recorded. Additionally, in the \textbf{FSI} and \textbf{FSD} simulations, two pressure sensors are placed within the fluid domain at $1~\text{mm}$ offset from the structural wall. Figure~\ref{fig:numerical_setup} shows the locations of these sensors. In all the simulations, the structure is subjected to the explosion of $250~\text{g}$ TNT at its center. The simulations are performed for a maximum time of $2$ ms. Beyond this time point, the containment structure is mainly vibrating about its new dynamic equilibrium configuration.
	
	%

	\subsection{Fluid-structure coupled simulation}\label{sec:fsi_results}

	
	In the fluid-structure coupled simulation (i.e.,~\textbf{FSI}), we solve the model equations presented in Section~\ref{sec:blast_mitigation} (Stage 3). The dynamics of the detonation induced fluid flow is computed in 2D using the cylindrical symmetry, while the mechanical response of the containment structure is calculated in 3D. The fluid inside the containment structure is initialized using the solution at the end of Stage 2, featuring a sharp interface between burnt gas and air. As a reference, Figure~\ref{fig:density_snapshots} shows the density and pressure solutions for a similar case with $226.8~\text{g}$ explosive. In Stage 2, burnt gas and air are modeled using different EOS, namely JWL and perfect gas. In Stage 3, we apply the JWL EOS to model both materials inside the containment structure, using the parameter values for the burnt gas. As mentioned at the end of  Section~\ref{sec:gas_expansion}, this simplification can be justified by the fact that for low density gases, JWL degenerates to perfect gas with Gr\"uneisen parameter $\gamma - 1 = \omega$.
	
\begin{figure}[!htb]
		\centering\includegraphics[width=120mm]{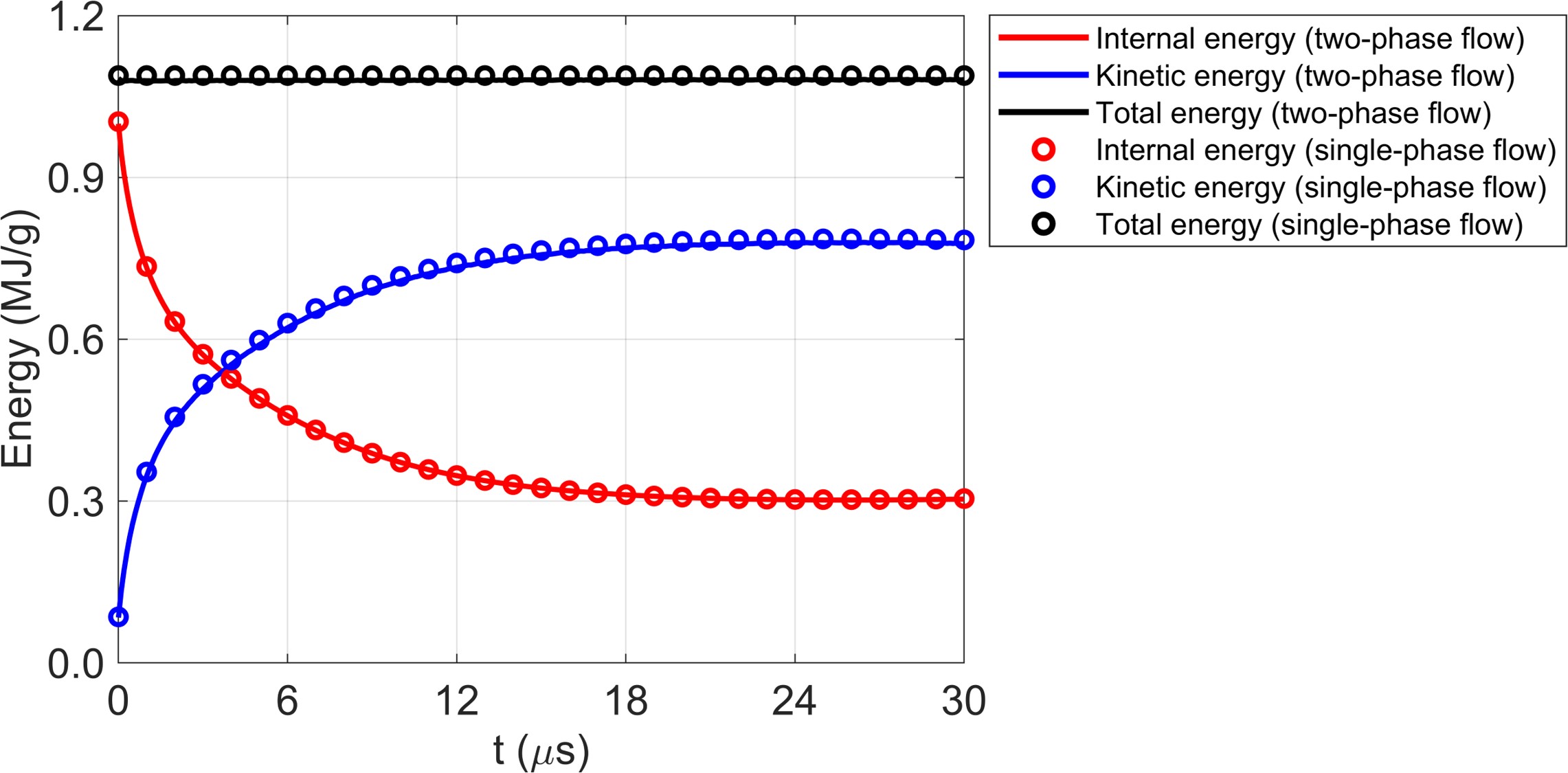}
		\caption{Comparison between the two-phase flow model in Stage 2 and the simplified single phase model used for Stage 3.}
		\label{fig:energy_integrals}
	\end{figure}

	To assess the effect of this simplification, we conduct a numerical experiment in which Stage 2 is repeated with the JWL EOS applied to both burnt gas and air. Figure~\ref{fig:energy_integrals} depicts the integrated kinetic, internal, and total energy obtained from the original and simplified models. In both cases, as the burnt gas expands, internal energy is converted into kinetic energy, while the total energy remains constant. The rate of conversion gradually decreases, and after about $30~\mu\text{s}$, a steady state is reached. The discrepancy between the two models is less than $1.5\%$.
	
	The ambient air outside the containment structure is still modeled using the perfect gas EOS, with $\gamma = 1.4$. Therefore, the fluid domain features two material subdomains separated by a moving fluid-structure interface.

	\subsubsection{Mesh sensitivity analysis}\label{sec:mesh_con_fsi}
	
	The fluid-structure coupled simulation is conducted on six pairs of meshes with varying resolutions, as outlined in Table~\ref{tab:mesh_resolution}. The structural element size is reduced by a factor of $25$, while the fluid element size is decreased by a factor of $50$. As the element size decreases, the time step size is also reduced accordingly.
	
	\begin{table}[!htb]
		\centering
		\begin{tabular}{cccccc}
			\toprule 
			~ & \multirow{3}{1.5cm}{\centering Time step size}
			& \multicolumn{2}{c}{Solid Mesh}                  
			& \multicolumn{2}{c}{Fluid Mesh} \\ \cmidrule{3-6}
			~ & 
			& \multirow{2}{*}{Number of nodes} 
			& \multirow{2}{*}{Element size (mm)} 
			& \multirow{2}{*}{Number of nodes} 
			& \multirow{2}{*}{Element size (mm)} \\
			&           &           &           &             &        \\ \cmidrule{3-6}
			Mesh pair $1$ & $5\times10^{-8}$  &   $119$    &   $25.0$   &    $1024$     &   $25.0$  \\
			Mesh pair $2$ & $5\times10^{-8}$  &   $405$    &   $12.5$   &    $2601$    &   $12.5$  \\
			Mesh pair $3$ & $5\times10^{-8}$  &   $2190$   &   $5.0$    &    $15625$    &   $5.0$   \\
			Mesh pair $4$ & $2\times10^{-8}$  &   $8416$   &   $2.5$    &    $54289$   &   $2.5$   \\
			Mesh pair $5$ & $2\times10^{-8}$  &   $8416$   &   $2.5$    &    $614656$   &   $0.5$   \\
			Mesh pair $6$ & $2.5\times10^{-9}$  &   $51520$  &   $1.0$    &    $203401$  &   $1.0$   \\
			\bottomrule
		\end{tabular}
		\caption{Spatial and temporal resolutions considered in the mesh sensitivity analysis.}
		\label{tab:mesh_resolution}
	\end{table}
	
\begin{figure}[!htb]
		\centering\includegraphics[width=150mm]{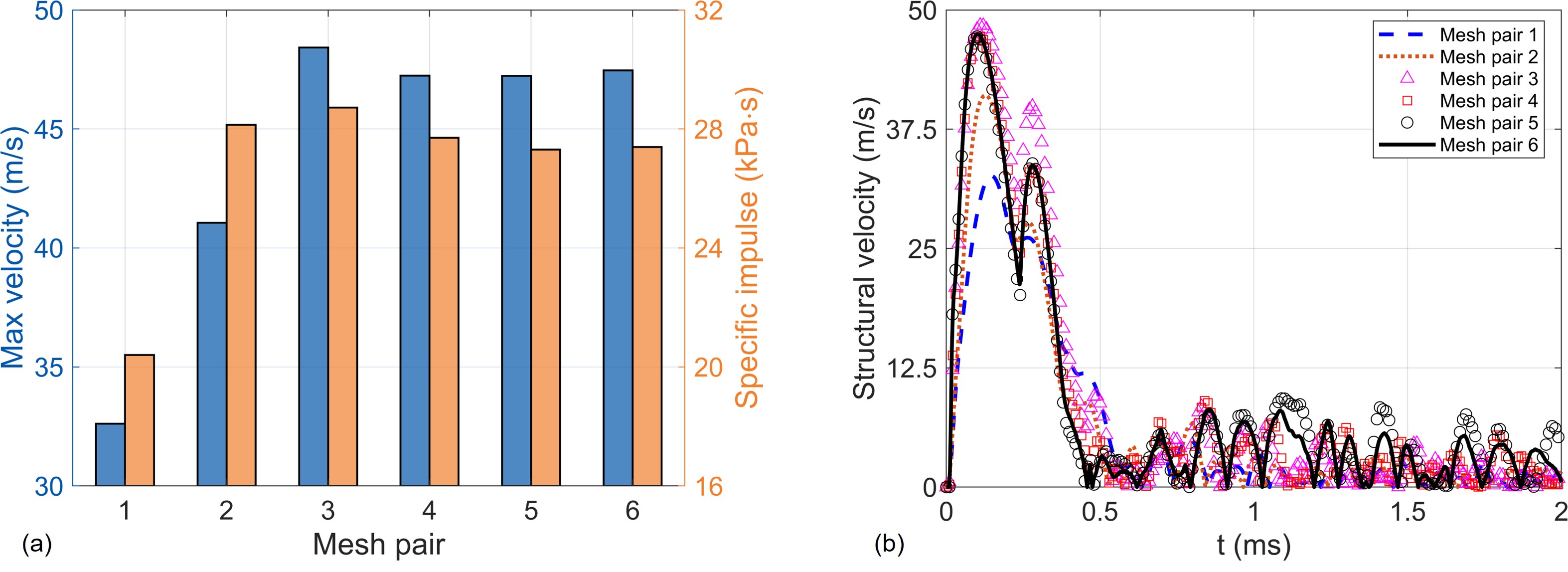}
		\caption{Comparison of results obtained with six pairs of meshes. (a) Maximum structural velocity and fluid pressure impulse at sensor 1. (b) Time-history of structural velocity magnitude at sensor 1.}
		\label{fig:fsi_convergence}
	\end{figure}

	Figure~\ref{fig:fsi_convergence} compares the results obtained from all six simulations. Subfigure (a) shows the maximum velocity at structural sensor $1$ and the specific pressure impulse at fluid sensor $1$ (integrated until $2~\text{ms}$). For both quantities, the solutions are no longer sensitive to mesh resolution beyond mesh pair $4$. Subfigure (b) compares the time-history of structural velocity at sensor 1. As expected, numerical error accumulates over time. The results from mesh pairs $4$, $5$, and $6$ match well up to approximately $1~\text{ms}$, during which four to five shock pulses have elapsed. After this point, the structural dynamics is dominated by elastic vibrations, a behavior captured by all mesh pairs, though quantitative discrepancies gradually increase with time. 
	
	The results shown in the remainder of this section are obtained with mesh pair $5$.

	\FloatBarrier

	\subsubsection{Results and discussion}\label{sec:coupling_results}
	
\begin{figure}[!htb]
		\centering\includegraphics[width=150mm]{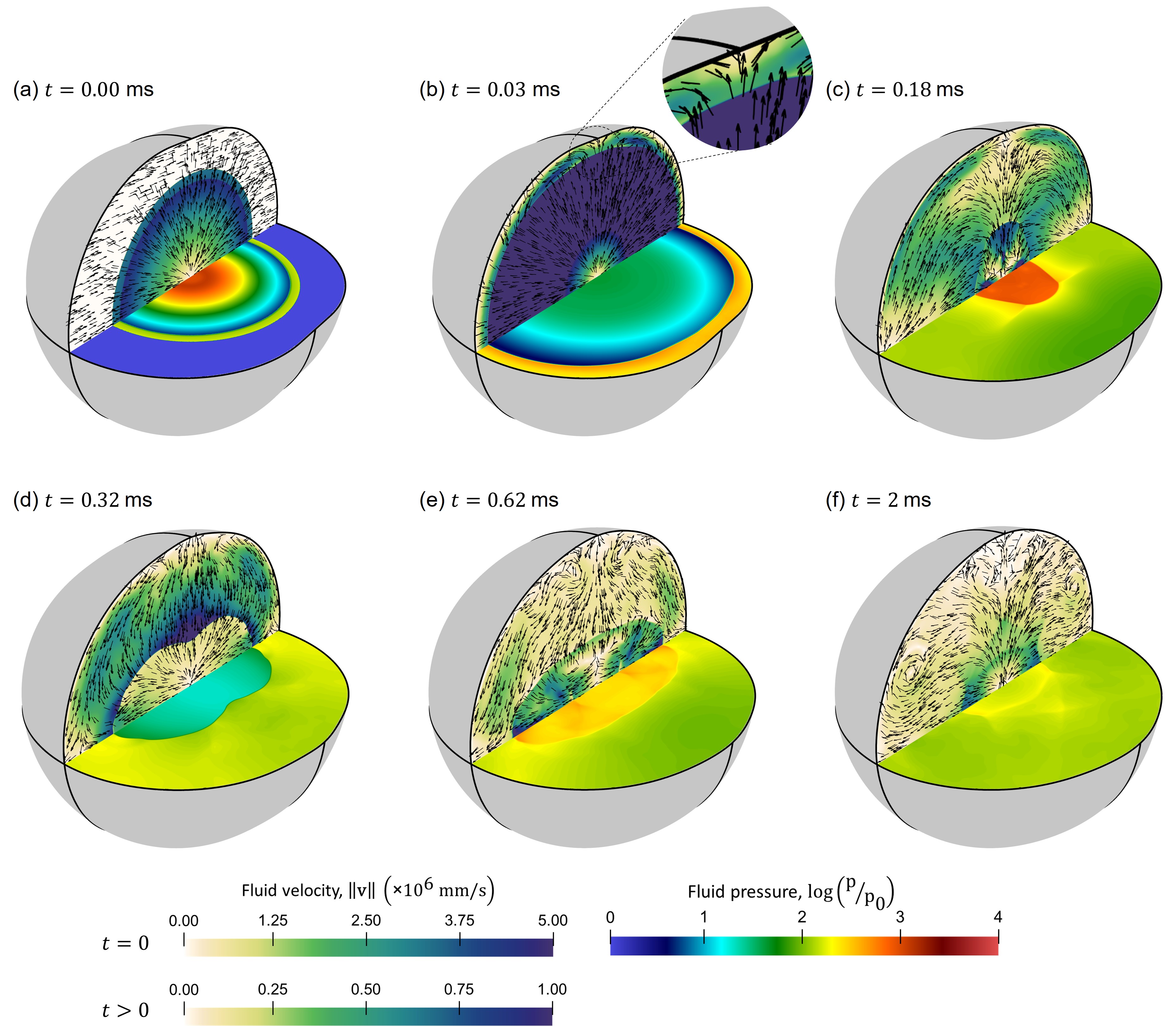}
		\caption{Snapshots of fluid velocity and pressure fields obtained from the \textbf{FSI} simulation. The velocity vectors are also shown as black arrows.}
		\label{fig:fluid_dynamics_overview}
	\end{figure}

	Figure~\ref{fig:fluid_dynamics_overview} provides an overview of the simulation results, showing the fluid pressure and velocity fields at six time instants together with the deformed structure. The fluid dynamics features complex wave propagation, reflections, and interactions. The structure is subjected to repeated impulsive loading from the reflected shock waves. Due to the structure's geometry, the reflected shock waves do not maintain the spherical shape. As a result, different points on the structural wall experience shock loads at different time instants with varying magnitudes. The structure undergoes plastic deformation. The cylindrical region deforms the most, and as a result, the structure's length-to-diameter ratio decreases from $1.125$ at time $t = 0$ to $1.07$ by the end of the simulation at $t=2~\text{ms}$.
	
\begin{figure}[!htb]
		\centering\includegraphics[width=150mm]{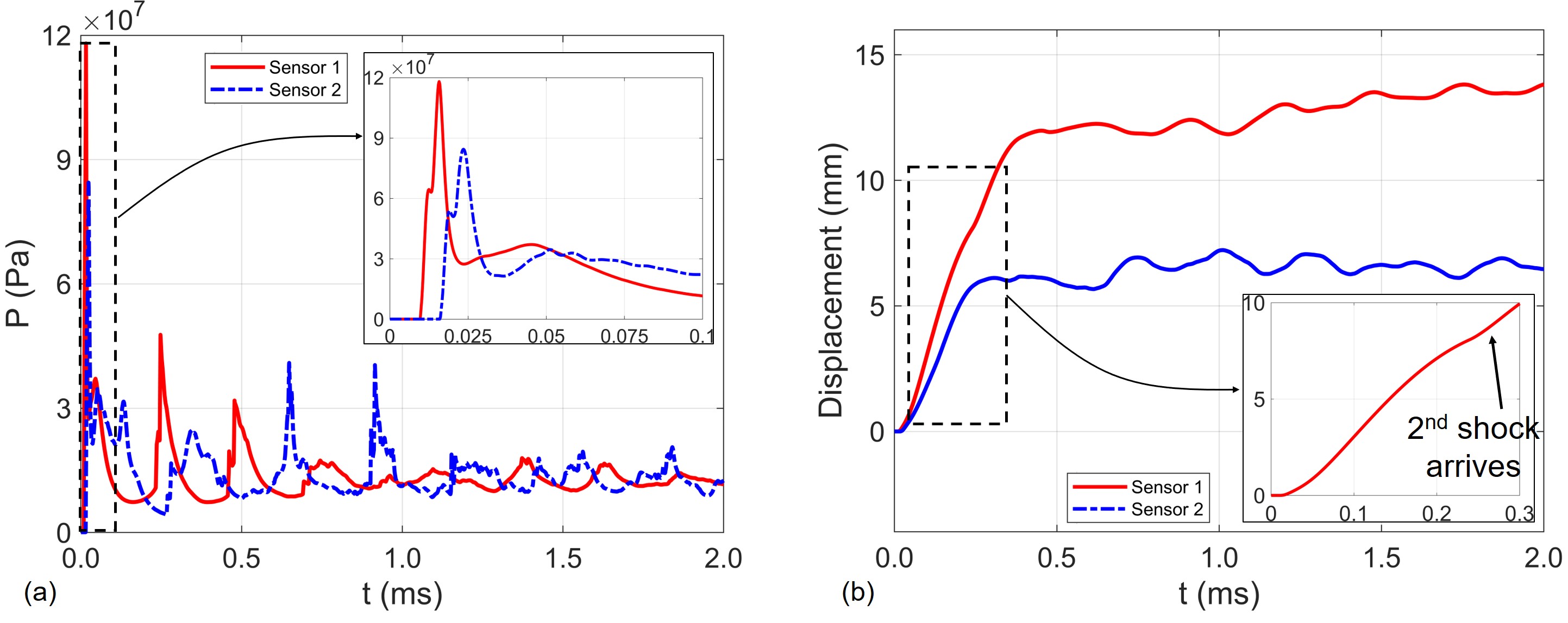}
		\caption{Temporal evolution of  (a) fluid pressure and (b) structural displacement, recorded at respective sensor locations shown in Figure~\ref{fig:numerical_setup}.}
		\label{fig:coupled_simulation_probe_plots}
	\end{figure}

	Figure~\ref{fig:coupled_simulation_probe_plots} presents the pressure and displacement data recorded by the fluid and structural sensors, respectively. Due to the non-spherical geometry of the containment structure, the results vary significantly between different sensor locations. At both fluid sensors, multiple shock pulses are captured. While the peak pressure of the initial pulse is considerably higher than that of the subsequent ones, its impulse is only marginally greater. The displacement data obtained from both sensors exhibit two phases: a rapid initial growth phase, followed by a combination of low-magnitude plastic deformations and elastic vibrations. The effect of repeated shock impacts is particularly clear at sensor $1$, where the first phase, lasting approximately $0.5~\text{ms}$, captures the first three shock pulses.
	
\begin{figure}[!htb]
		\centering\includegraphics[width=150mm]{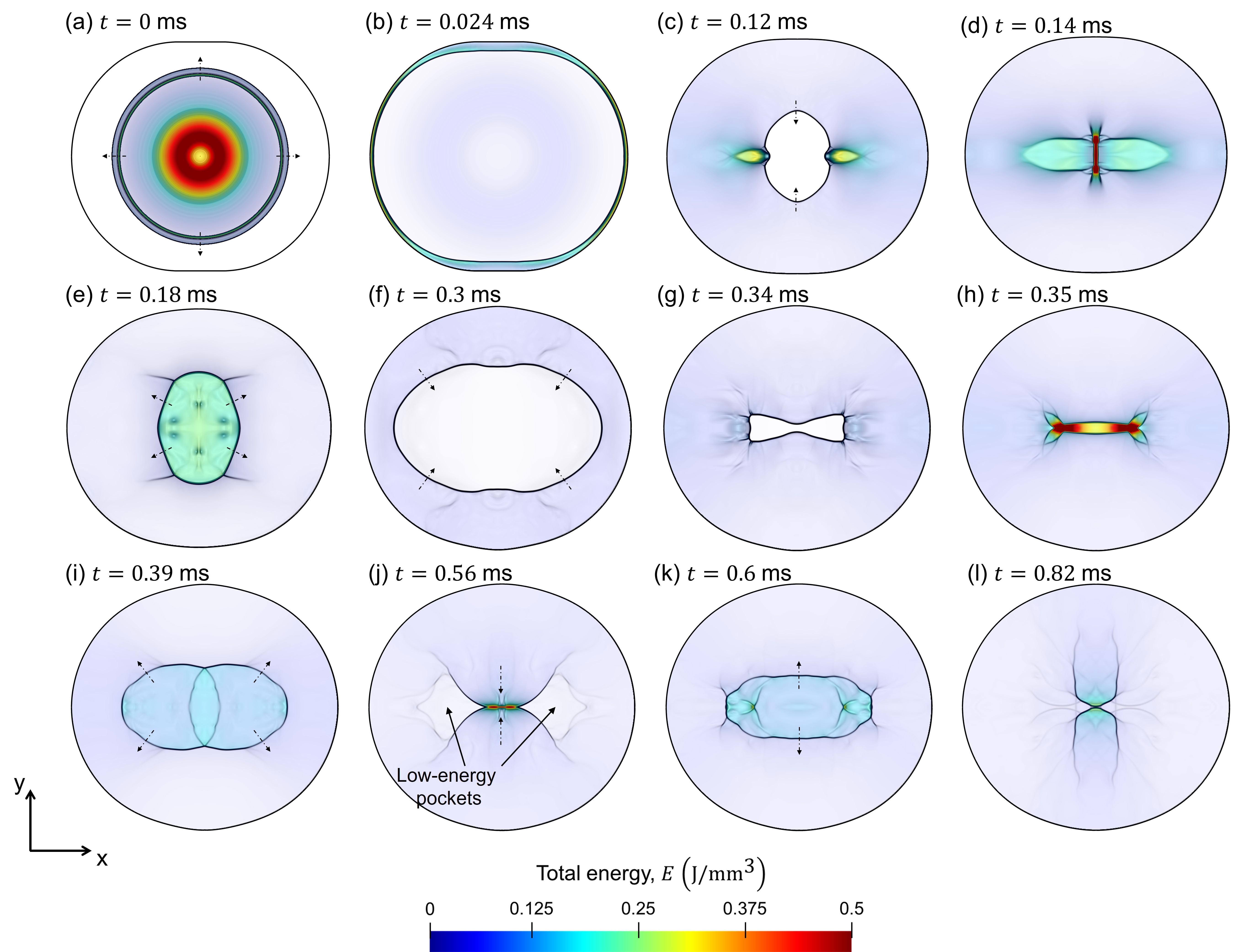}
		\caption{Total (i.e., internal and kinetic) energy per unit volume within the fluid domain, overlaid with its gradient (magnitude) to track wave propagation.}
		\label{fig:simulation_snaps_fluid}
	\end{figure}

\begin{figure}[!htb]
		\centering\includegraphics[width=150mm]{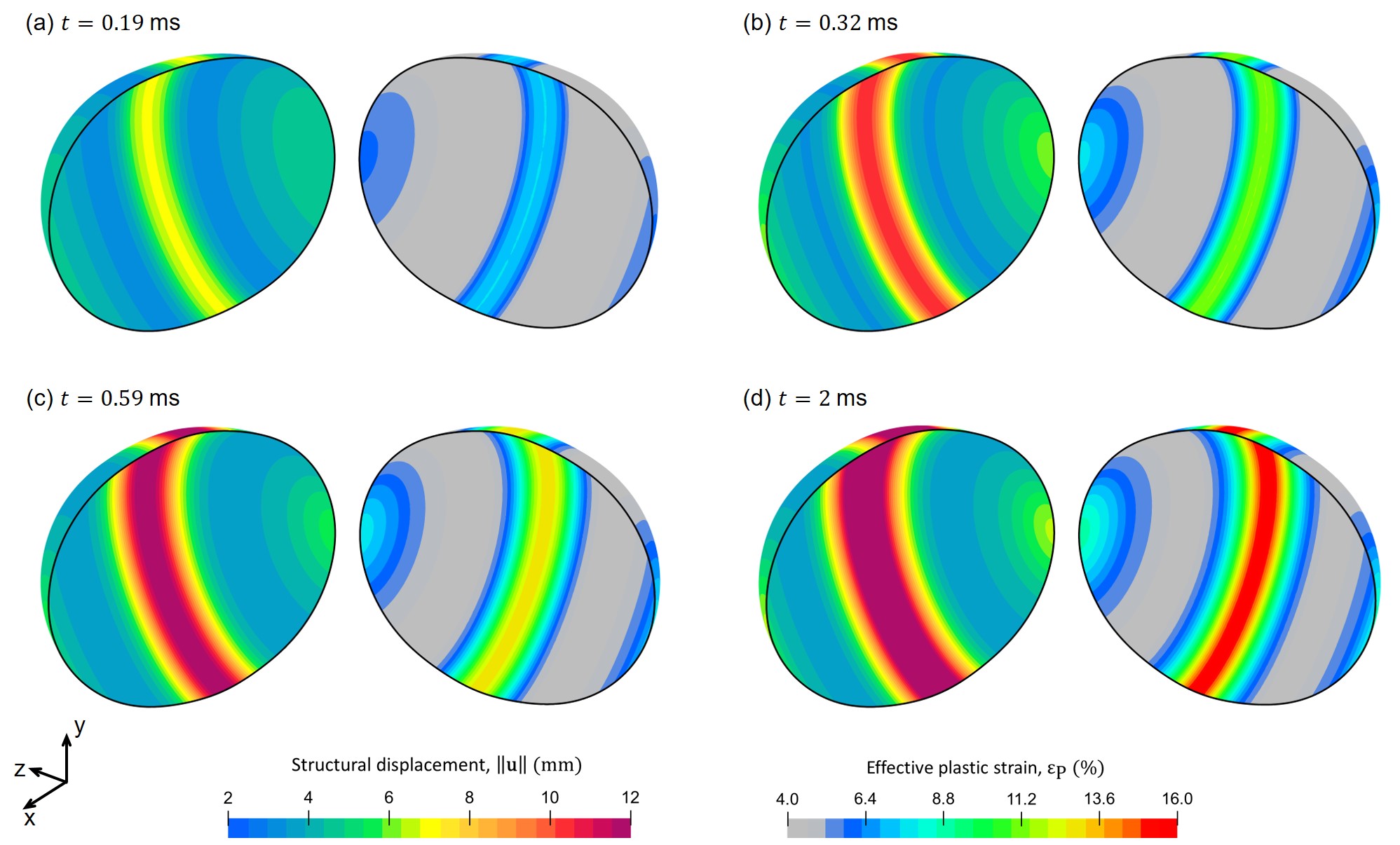}
		\caption{Structural response to the internal explosion. In each subfigure, the displacement field is visualized on the left, while the effective plastic strain is shown on the right.}
		\label{fig:simulation_snaps_solid}
	\end{figure}

	To describe the fluid and structural dynamics in detail, we also visualize the fluid's total energy in Figure~\ref{fig:simulation_snaps_fluid}, and the structure's displacement and plastic strain in Figure~\ref{fig:simulation_snaps_solid}. As expected, the first shock impact occurs on the mid-plane of the containment structure (i.e., $x=0$), as it is closest to the point of detonation. This leads to high stress along the mid-plane, causing the material to yield (Figure~\ref{fig:simulation_snaps_solid}(a)). By $t = 0.024~\text{ms}$, the incident shock wave has reached the entire structural wall, and the reflected waves are converging inward from all directions. The two ellipsoidal endpoints of the structure have also started to deform plastically, as they are also impacted by the shock wave in the normal direction (Figure~\ref{fig:simulation_snaps_solid}(a)).
	
	It is notable that the reflected shock wave does not propagate at a uniform speed. In particular, the reflections from the mid-plane and nearby regions travel slower than those from the ellipsoidal ends. This variation is due to both structural geometry and fluid-structure interaction. As the mid-plane region expands outward, it drags the internal fluid with it, thereby slowing down the reflected wave. This phenomenon can be seen in the velocity field of Figure~\ref{fig:fluid_dynamics_overview}(b). The ellipsoidal ends, although also moving outward, do so at a lower speed. More importantly, the shock waves reflected from the ellipsoidal regions converge along the center-line (i.e., $y=z=0$), forming a stronger and faster shock moving inward along the axial (i.e., $x$) direction. This behavior is captured by Figure~\ref{fig:simulation_snaps_fluid}(c). As a result, although the structure's axial length exceeds its diameter, the reflected shock waves converge on its mid-plane ($x=0$), as shown in Figure~\ref{fig:simulation_snaps_fluid}(d).
	
	The reflected shock waves impact the structure between $0.23~\text{ms}$ and $0.25~\text{ms}$. Again, the mid-plane region is impacted first. Although the peak pressure from the second shock pulse ($4.77\times10^7~\text{Pa}$) is much lower than that of the incident shock ($11.8\times10^7~\text{Pa}$), the impulses of both pulses are comparable: $3.89\times10^3~\text{Pa}\cdot\text{s}$ for the first pulse and $3.12\times10^3~\text{Pa}\cdot\text{s}$ for the second. This indicates that the second shock pulse has a clear effect on the structure. Figures~\ref{fig:simulation_snaps_solid}(b) demonstrate that this pulse is strong enough to induce yielding in the structural material, increasing mid-plane deformations from $8.1~\text{mm}$ to $11.9~\text{mm}$, with a corresponding rise in plastic strain from $8.8\%$ to $12.7\%$. Moreover, Figure~\ref{fig:coupled_simulation_probe_plots}(b) shows that after the first pulse elapses (around $0.2~\text{ms}$), the structural velocity starts to decrease. However, when the second shock pulse arrives, the velocity begins to increase once again. 
	
	Due to the shape of the second shock pulse, as shown in Figure~\ref{fig:simulation_snaps_fluid}(e),  the initial impact points shift away from the mid-plane. Again, the reflected shocks do not propagate at a uniform speed. The shock front forms a dented ellipsoidal shape, as shown in Figure~\ref{fig:simulation_snaps_fluid}(f). The waves reflected from the ellipsoidal ends once again converge on the center-line (Figure~\ref{fig:simulation_snaps_fluid}(g)). By this time, the waves reflected from the cylindrical region have also reached the center-line. This coincidence creates two high-energy zones along the center-line, acting like two point sources emitting spherically expanding waves (Figure~\ref{fig:simulation_snaps_fluid}(h)). The two waves collide on the mid-plane, creating two secondary shocks that follow the main ones (Figure~\ref{fig:simulation_snaps_fluid}(i)). 
	
	When the two main shocks reflect from the structural wall, they collide with the secondary shocks and get decelerated. This leads to two low-energy zones as shown in Figure~\ref{fig:simulation_snaps_fluid}(j). When the inward-propagating reflected wave from the mid-plane converges, it generates another outward-moving shock wave. This shock propagates through the low-energy zones along the axial direction while encountering resistance in the radial direction (i.e., $y$). As a result, the shock retains much of its energy and arrives first at the ellipsoidal ends of the containment structure, leading to a spike in pressure recorded by fluid sensor 2 between $0.6~\text{ms}$ and $1~\text{ms}$, shown in Figure~\ref{fig:coupled_simulation_probe_plots}(a).
	
	In summary, the containment structure considered in this case is not able to elastically withstand the explosion loads. It dissipates the energy of the explosion mostly by accumulating plastic strain. Yielding begins immediately after the arrival of the initial shock, with plastic strain continuing to grow during subsequent shock pulses.  By the end of the simulation, the maximum effective plastic strain is found to be $16.1\%$ (Figure~\ref{fig:simulation_snaps_solid}(d)), nearing the material's failure threshold. After several reverberations, the shock waves gradually weaken as the internal volume of the structure expands by approximately $30\%$.

	\subsection{Comparison with decoupled simulations}\label{sec:fluid_estimates}
	
\begin{figure}[!htb]
		\centering\includegraphics[width=120mm]{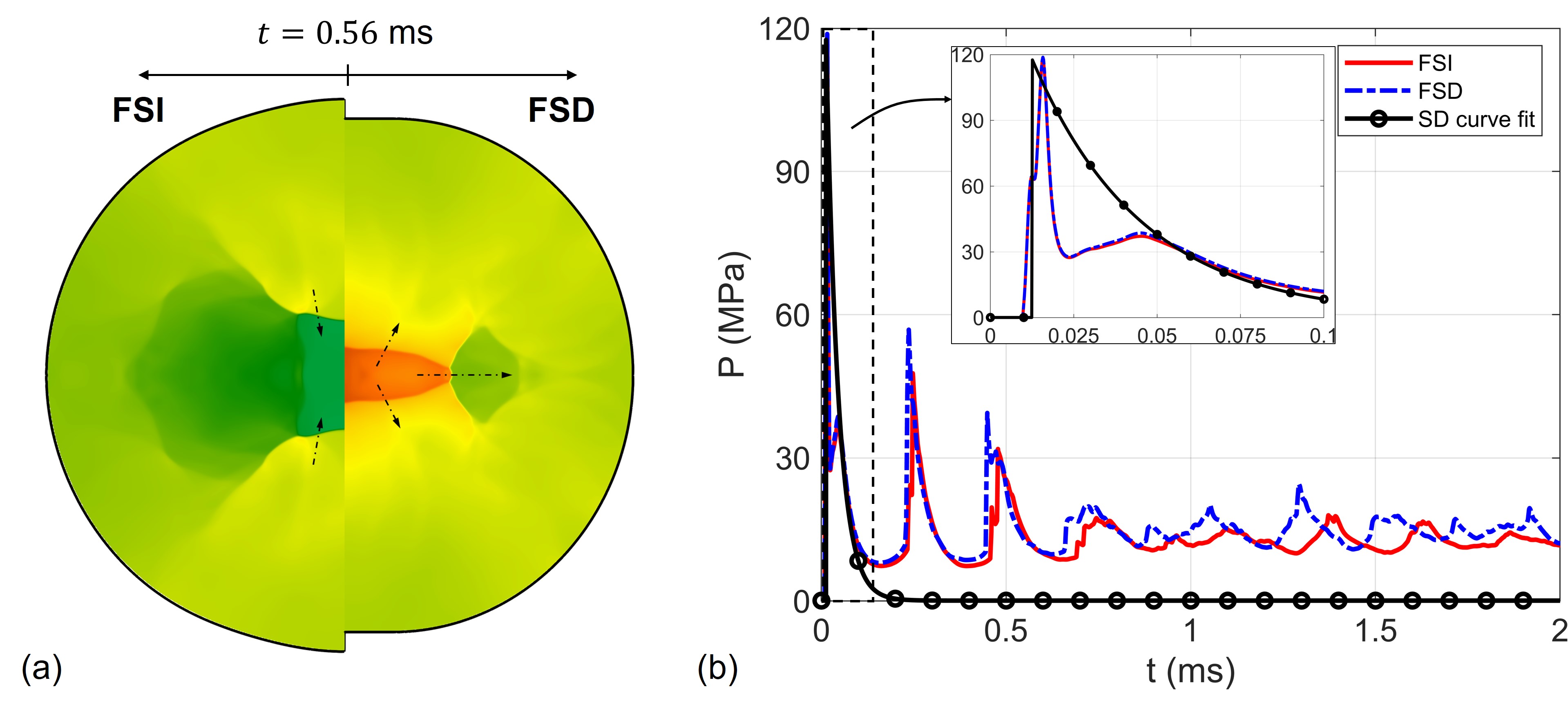}
		\caption{Comparison of fluid pressure in \textbf{FSI}, \textbf{FSD} (a), and the curve fit used in \textbf{SD} (b).}
		\label{fig:couple_decouple}
	\end{figure}

	
	We compare the \textbf{FSI} simulation with two simplified methods, \textbf{FSD} and \textbf{SD}, in which the structural dynamics is decoupled from that of the fluid. In \textbf{FSD}, a fluid dynamics simulation is performed first, with the containment structure represented by a fixed wall boundary. Figure~\ref{fig:couple_decouple} compares the fluid pressure results obtained from \textbf{FSI} and \textbf{FSD}. It can be observed that \textbf{FSD} is able to capture the first shock pulse accurately, with the error in peak pressure less than $1\%$.  This can be attributed to the high acoustic impedance of the steel material ($4.726\times10^{4}~\text{Pa}\cdot\text{s}/\text{m}$) compared to the burnt gas ($7.513 ~\text{Pa}\cdot\text{s}/\text{m}$), causing the majority of the energy to be reflected back into the fluid rather than transmitted through the material.
	
	However, after the first shock wave elapses, the \textbf{FSD} result begins to deviate from that of \textbf{FSI}, primarily due to the fluid simulation in \textbf{FSD} neglecting transient structural deformations. Because the structure is actually expanding, \textbf{FSD} overestimates the pressure magnitude, and the error increases in time. The peak pressure values captured during the second and third shock pulses differ from the \textbf{FSI} results by $16.06\%$ and $50.23\%$, respectively.
	
	The phase difference between the two simulations also increases over time. From the second shock pulse onward, the pulses from \textbf{FSD} arrive earlier than their counterparts in \textbf{FSI}. This discrepancy arises because in the fluid simulation of \textbf{FSD}, the structure remains fixed, resulting in a more energetic fluid flow and higher shock propagation speeds. Additionally, by neglecting structural expansion, \textbf{FSD} shortens the travel distance for reflected shock waves, further accelerating their arrival. After $1~\text{ms}$, the results of the two simulations are completely different.
	
	In the \textbf{SD} simulation, the fluid flow is not simulated.  Instead, we directly apply a transient pressure load on the structural model, using the Friedlander equation \eqref{eq:friedlander}. We fit this exponentially decaying function to the first shock pulse recorded at fluid sensor 1 during the \textbf{FSI} simulation, which yields $p_a=100~\text{kPa}$, $p_{r,max}=118.05~\text{MPa}$, $t_a=12.47~\mu\text{s}$, $t_{d+}=2~\text{ms}$, and $b=59.54$. The resulting pressure load is plotted in Figure~\ref{fig:couple_decouple}(b). Unlike in \textbf{FSD}, spatial variations in pressure are ignored, and the same pressure load is applied across the entire structural wall. Therefore, the effect of the structural geometry on pressure loading is lost.
	
\begin{figure}[!htb]
		\centering\includegraphics[width=120mm]{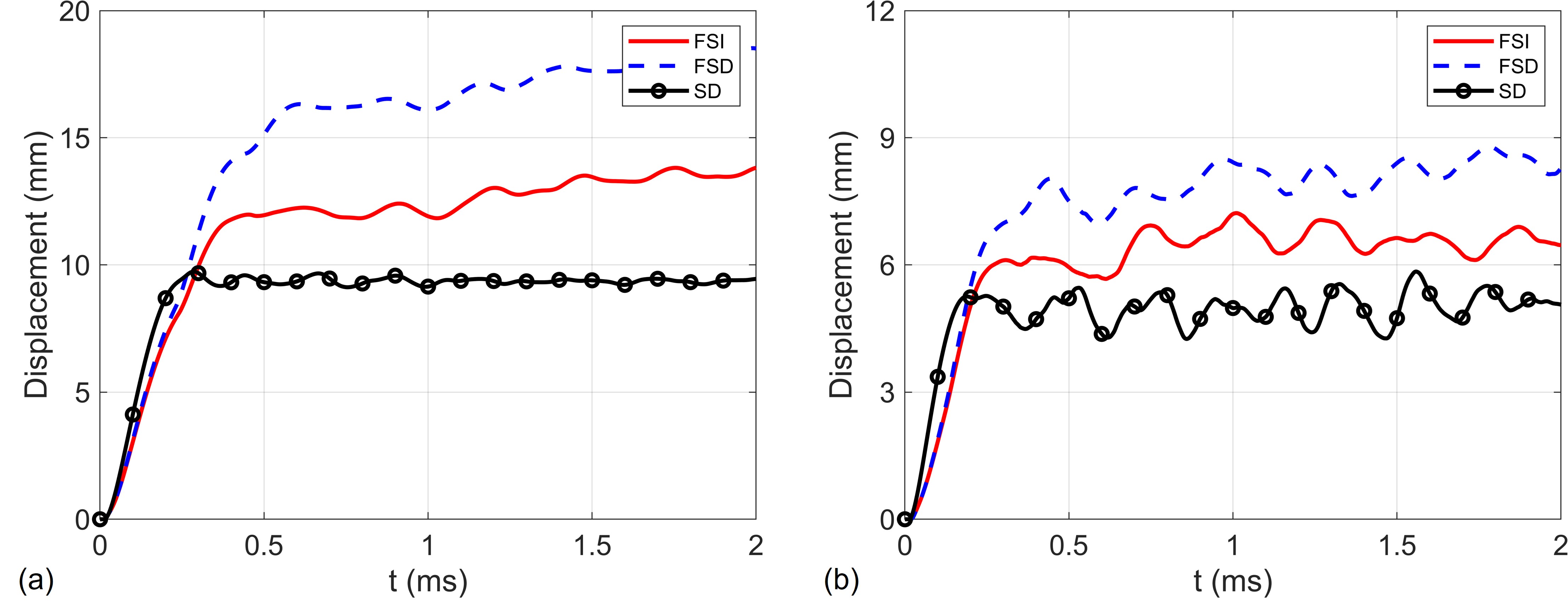}
		\caption{Comparison of structural displacement at sensor locations. (a) Sensor 1. (b) Sensor 2.}
		\label{fig:displacement_comparison}
	\end{figure}
\begin{figure}[!htb]
		\centering\includegraphics[width=120mm]{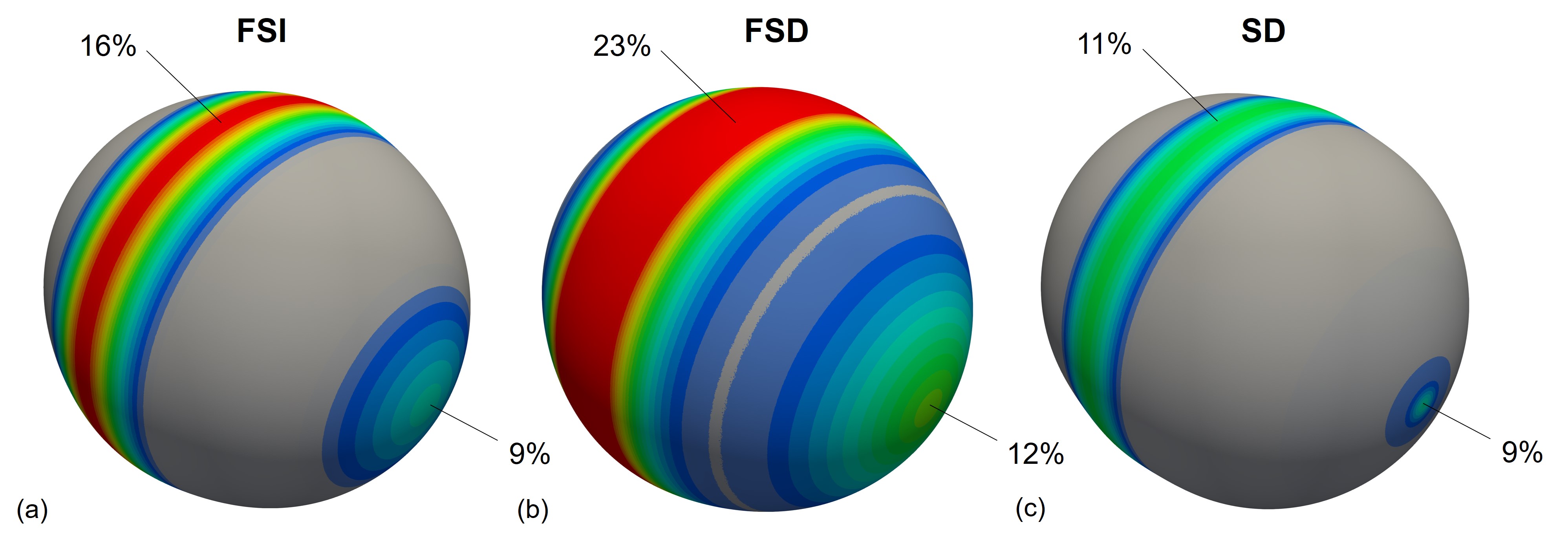}
		\caption{Comparison of effective plastic strain at the end of each simulation ($t=2~\text{ms}$).}
		\label{fig:final_plastic_strains}
	\end{figure}

	Figures~\ref{fig:displacement_comparison} and~\ref{fig:final_plastic_strains} compare the structural results obtained from the three simulations. At sensor $1$, \textbf{FSD} overpredicts the maximum structural displacement by $30.18\%$, as the structure is subjected to an overestimated pressure load. For the same reason, it overpredicts the effective plastic strain by $43.75\%$. The \textbf{SD} simulation underpredicts both quantities, as the pressure load only captures the first shock pulse but neglects the subsequent ones. The errors in maximum displacement and effective plastic strain are found to be $23.03\%$ and $31.25\%$, respectively.
	
	The findings presented in this section align with those of Aune {\it et al.}, who conducted experimental and numerical studies on the effects of fluid-structure coupling on thin deformable steel plates subjected to shock-generated pressure in a shock tube \cite{aune_shock_2016, aune_fluid_2021}. The shock tube, a partially confined structure, features an open end where the shock originates and a closed end where the thin plate resides. In this setup, the resulting fluid pressure consists of a single peak followed by an exponential decay as the reflected shock receded toward the open end. Despite this, incorporating the fluid-structure coupling in their simulations reduced the mid-point deflections by 4–14\%, depending on the shock intensity \cite{aune_fluid_2021}. For fully confined containment structures, such as the one considered here, this coupling becomes even more critical due to the presence of complex wave reflections and interactions, as is evident from the findings presented in this section.

	\section{Conclusion}\label{sec:conclusion}
	
	We have developed a computational model to predict the dynamic response of lightweight structures subjected to internal explosions, allowing for large plastic deformations of the structure. The analysis starts with the chemical reaction of explosion and accounts for the interaction between the structure and the shock-dominated flow of gaseous explosion products. To address varying length and time scales of different physical processes, and to optimize computational efficiency through symmetry, the simulation is divided into three stages. The governing fluid and structural equations are solved separately by finite volume and finite element methods. A few special techniques, including embedded boundary, FIVER, and level set methods, are utilized to track material interfaces and enforce interface conditions. These novel methods distinguish the proposed computational model from existing methods implemented in commercial solvers like LS-Dyna (e.g., CESE and ALE methods). Several verification tests, including mesh convergence and simplified model analyses, were performed to assess the accuracy of the solver.
	
	The computational model has been implemented in open-source fluid and structural dynamics solvers, M2C~\cite{m2c} and Aero-S~\cite{aeros}. Its effectiveness is demonstrated through the case study of a thin-walled steel chamber subjected to the detonation of $250$ g of TNT. In this case, the chamber undergoes plastic deformation, with internal volume increasing by $30\%$. However, the maximum strain remains below the material's fracture limit. The importance of fluid-structure interaction is evident in several findings.
	
	\begin{itemize}
		
		\item Although the initial shock pulse carries the highest peak pressure, subsequent pulses from wave reflections also contribute to structural deformation. While the peak pressure declines rapidly, the impulses of the first three shock pulses are comparable, causing plastic deformation to increase and broaden over these pulses. Therefore, a model using only the initial shock pulse to estimate the load would significantly underpredict the maximum displacement and strain.
		
		\item Complex shock interactions arise not only from the geometry of the chamber but also from the nature of the fluid flow. The high energy carried by the burnt gas makes the flow highly compressible, causing shock waves to propagate at variable speeds. Neglecting these variations in shock speeds --- such as by using linear acoustic theory --- could lead to substantial errors when calculating the pressure loads on the chamber wall.
		
		\item The structural dynamics also reciprocally influences the fluid flow. The chamber's expansion increases the travel distance for shock reflections and reduces the energy density of the flow. Treating the chamber as a fixed boundary while simulating the internal fluid dynamics would overpredict the magnitude of shock pulses and underpredicts the time intervals between successive pulses.
	\end{itemize}
	
	Overall, the results suggest that allowing permanent deformations in single-use explosion containment chambers offers potential for weight reduction. They also underscore the importance of accounting for fluid-structure interaction effects in the design of this type of chambers. Relying on loads derived from decoupled fluid analyses may lead to overly conservative and heavier designs for such containment structures, a concern that also applies to partially confined structures (e.g., \cite{aune_shock_2016, aune_fluid_2021}). Several simplifications made in this study can be refined in future research. For example, the Chapman-Jouguet theory can be replaced with more detailed explosion models that resolve the chemical reaction zone. Additionally, the structural model used in the case study can be extended to accommodate more complex structural geometry and materials (e.g., composite materials).

	\section*{Declaration of Competing Interest}
	
	The authors declare that they have no known competing financial
	interests or personal relationships that could have appeared
	to influence the work reported in this paper.

	\section*{Acknowledgment}
	A.N, S.I., and K.W. gratefully acknowledge the support of DynaSafe US LLC through a research contract to Virginia Tech, and the support of Office of Naval Research (ONR) under awards N00014-23-1-2447 and N00014-24-1-2509. A.N. and K.W. appreciate the insightful discussions with Mr.~Harley Heaton. X.S. gratefully acknowledges the support of the University of Kentucky through the faculty startup fund.

	\printcredits
	
	
	\newpage
	\setcitestyle{authoryear}
	\bibliographystyle{cas-model2-names}
	
	\bibliography{manuscript}

\begin{thebibliography}{44}
\expandafter\ifx\csname natexlab\endcsname\relax\def\natexlab#1{#1}\fi
\providecommand{\url}[1]{\texttt{#1}}
\providecommand{\href}[2]{#2}
\providecommand{\path}[1]{#1}
\providecommand{\DOIprefix}{doi:}
\providecommand{\ArXivprefix}{arXiv:}
\providecommand{\URLprefix}{URL: }
\providecommand{\Pubmedprefix}{pmid:}
\providecommand{\doi}[1]{\href{http://dx.doi.org/#1}{\path{#1}}}
\providecommand{\Pubmed}[1]{\href{pmid:#1}{\path{#1}}}
\providecommand{\bibinfo}[2]{#2}
\ifx\xfnm\relax \def\xfnm[#1]{\unskip,\space#1}\fi
\bibitem[{AeroF(2023)}]{aerof}
\bibinfo{author}{AeroF}, \bibinfo{year}{2023}.
\newblock \URLprefix \url{https://bitbucket.org/frg/aero-f/src/master/}.
\bibitem[{AeroS(2023)}]{aeros}
\bibinfo{author}{AeroS}, \bibinfo{year}{2023}.
\newblock \URLprefix \url{https://bitbucket.org/frg/aero-s/src/master/}.
\bibitem[{Alia and Souli(2006)}]{alia_high_2006}
\bibinfo{author}{Alia, A.}, \bibinfo{author}{Souli, M.}, \bibinfo{year}{2006}.
\newblock \bibinfo{title}{High explosive simulation using multi-material
  formulations}.
\newblock \bibinfo{journal}{Applied thermal engineering} \bibinfo{volume}{26},
  \bibinfo{pages}{1032--1042}.
\bibitem[{Army(1986)}]{army_fundamentals_1986}
\bibinfo{author}{Army, U.}, \bibinfo{year}{1986}.
\newblock \bibinfo{title}{Fundamentals of protective design for conventional
  weapons}.
\newblock \bibinfo{journal}{Technical manual TM} , \bibinfo{pages}{5--855}.
\bibitem[{Aune et~al.(2016)Aune, Fagerholt, Langseth and
  Børvik}]{aune_shock_2016}
\bibinfo{author}{Aune, V.}, \bibinfo{author}{Fagerholt, E.},
  \bibinfo{author}{Langseth, M.}, \bibinfo{author}{Børvik, T.},
  \bibinfo{year}{2016}.
\newblock \bibinfo{title}{A shock tube facility to generate blast loading on
  structures}.
\newblock \bibinfo{journal}{International Journal of Protective Structures}
  \bibinfo{volume}{7}, \bibinfo{pages}{340--366}.
\bibitem[{Aune et~al.(2021)Aune, Valsamos, Casadei, Langseth and
  B{\o}rvik}]{aune_fluid_2021}
\bibinfo{author}{Aune, V.}, \bibinfo{author}{Valsamos, G.},
  \bibinfo{author}{Casadei, F.}, \bibinfo{author}{Langseth, M.},
  \bibinfo{author}{B{\o}rvik, T.}, \bibinfo{year}{2021}.
\newblock \bibinfo{title}{Fluid-structure interaction effects during the
  dynamic response of clamped thin steel plates exposed to blast loading}.
\newblock \bibinfo{journal}{International Journal of Mechanical Sciences}
  \bibinfo{volume}{195}, \bibinfo{pages}{106263}.
\bibitem[{Baker(1960)}]{baker_elastic_1960}
\bibinfo{author}{Baker, W.E.}, \bibinfo{year}{1960}.
\newblock \bibinfo{title}{The elastic-plastic response of thin spherical shells
  to internal blast loading} .
\bibitem[{Belytschko et~al.(1984)Belytschko, Lin and
  Chen-Shyh}]{belytschko_explicit_1984}
\bibinfo{author}{Belytschko, T.}, \bibinfo{author}{Lin, J.I.},
  \bibinfo{author}{Chen-Shyh, T.}, \bibinfo{year}{1984}.
\newblock \bibinfo{title}{Explicit algorithms for the nonlinear dynamics of
  shells}.
\newblock \bibinfo{journal}{Computer Methods in Applied Mechanics and
  Engineering} \bibinfo{volume}{42}, \bibinfo{pages}{225--251}.
\bibitem[{Cao et~al.(2021)Cao, Wang, Coutier-Delgosha and
  Wang}]{cao_shock_2021}
\bibinfo{author}{Cao, S.}, \bibinfo{author}{Wang, G.},
  \bibinfo{author}{Coutier-Delgosha, O.}, \bibinfo{author}{Wang, K.},
  \bibinfo{year}{2021}.
\newblock \bibinfo{title}{Shock-induced bubble collapse near solid materials:
  effect of acoustic impedance}.
\newblock \bibinfo{journal}{Journal of Fluid Mechanics} \bibinfo{volume}{907}.
\newblock \DOIprefix\doi{10.1017/jfm.2020.810}.
\bibitem[{Cao et~al.(2019)Cao, Zhang, Liao, Zhong and Wang}]{cao_shock_2019}
\bibinfo{author}{Cao, S.}, \bibinfo{author}{Zhang, Y.}, \bibinfo{author}{Liao,
  D.}, \bibinfo{author}{Zhong, P.}, \bibinfo{author}{Wang, K.G.},
  \bibinfo{year}{2019}.
\newblock \bibinfo{title}{Shock-induced damage and dynamic fracture in
  cylindrical bodies submerged in liquid}.
\newblock \bibinfo{journal}{International journal of solids and structures}
  \bibinfo{volume}{169}, \bibinfo{pages}{55--71}.
\bibitem[{Chang(1995)}]{chang_method_1995}
\bibinfo{author}{Chang, S.C.}, \bibinfo{year}{1995}.
\newblock \bibinfo{title}{The method of space-time conservation element and
  solution element—a new approach for solving the navier-stokes and euler
  equations}.
\newblock \bibinfo{journal}{Journal of computational Physics}
  \bibinfo{volume}{119}, \bibinfo{pages}{295--324}.
\bibitem[{Dobratz(1981)}]{osti_6530310}
\bibinfo{author}{Dobratz, B.M.}, \bibinfo{year}{1981}.
\newblock \bibinfo{title}{Llnl explosives handbook: properties of chemical
  explosives and explosives and explosive simulants} \URLprefix
  \url{https://www.osti.gov/biblio/6530310}, \DOIprefix\doi{10.2172/6530310}.
\bibitem[{Dong et~al.(2010)Dong, Li and Zheng}]{dong_interactive_2010}
\bibinfo{author}{Dong, Q.}, \bibinfo{author}{Li, Q.M.}, \bibinfo{author}{Zheng,
  J.Y.}, \bibinfo{year}{2010}.
\newblock \bibinfo{title}{Interactive mechanisms between the internal blast
  loading and the dynamic elastic response of spherical containment vessels}.
\newblock \bibinfo{journal}{International journal of impact engineering}
  \bibinfo{volume}{37}, \bibinfo{pages}{349--358}.
\bibitem[{Duffey and Mitchell(1973)}]{duffey_containment_1973}
\bibinfo{author}{Duffey, T.}, \bibinfo{author}{Mitchell, D.},
  \bibinfo{year}{1973}.
\newblock \bibinfo{title}{Containment of explosions in cylindrical shells}.
\newblock \bibinfo{journal}{International Journal of Mechanical Sciences}
  \bibinfo{volume}{15}, \bibinfo{pages}{237--249}.
\bibitem[{Duffey et~al.(2002)Duffey, Rodriguez and
  Romero}]{duffey_detonation-induced_2002}
\bibinfo{author}{Duffey, T.A.}, \bibinfo{author}{Rodriguez, E.A.},
  \bibinfo{author}{Romero, C.}, \bibinfo{year}{2002}.
\newblock \bibinfo{title}{Detonation-induced dynamic pressure loading in
  containment vessels}.
\newblock \bibinfo{journal}{Rep. LA-UR-02-0366, Los Alamos National Laboratory,
  Los Alamos, NM} .
\bibitem[{Duffey and Romero(2003)}]{duffey_strain_2003}
\bibinfo{author}{Duffey, T.A.}, \bibinfo{author}{Romero, C.},
  \bibinfo{year}{2003}.
\newblock \bibinfo{title}{Strain growth in spherical explosive chambers
  subjected to internal blast loading}.
\newblock \bibinfo{journal}{International Journal of Impact Engineering}
  \bibinfo{volume}{28}, \bibinfo{pages}{967--983}.
\bibitem[{Farhat et~al.(2012)Farhat, Gerbeau and Rallu}]{farhat_fiver_2012}
\bibinfo{author}{Farhat, C.}, \bibinfo{author}{Gerbeau, J.F.},
  \bibinfo{author}{Rallu, A.}, \bibinfo{year}{2012}.
\newblock \bibinfo{title}{Fiver: A finite volume method based on exact
  two-phase riemann problems and sparse grids for multi-material flows with
  large density jumps}.
\newblock \bibinfo{journal}{Journal of Computational Physics}
  \bibinfo{volume}{231}, \bibinfo{pages}{6360--6379}.
\bibitem[{Farhat et~al.(2010)Farhat, Rallu, Wang and
  Belytschko}]{farhat_robust_2010}
\bibinfo{author}{Farhat, C.}, \bibinfo{author}{Rallu, A.},
  \bibinfo{author}{Wang, K.}, \bibinfo{author}{Belytschko, T.},
  \bibinfo{year}{2010}.
\newblock \bibinfo{title}{Robust and provably second-order explicit-explicit
  and implicit-explicit staggered time-integrators for highly non-linear
  compressible fluid-structure interaction problems}.
\newblock \bibinfo{journal}{International Journal for Numerical Methods in
  Engineering} \bibinfo{volume}{84}, \bibinfo{pages}{73--107}.
\newblock \DOIprefix\doi{10.1002/nme.2883}.
\bibitem[{Farhat et~al.(2013)Farhat, Wang, Main, Kyriakides, Lee, Ravi-Chandar
  and Belytschko}]{farhat_dynamic_2013}
\bibinfo{author}{Farhat, C.}, \bibinfo{author}{Wang, K.},
  \bibinfo{author}{Main, A.}, \bibinfo{author}{Kyriakides, S.},
  \bibinfo{author}{Lee, L.H.}, \bibinfo{author}{Ravi-Chandar, K.},
  \bibinfo{author}{Belytschko, T.}, \bibinfo{year}{2013}.
\newblock \bibinfo{title}{Dynamic implosion of underwater cylindrical shells:
  experiments and computations}.
\newblock \bibinfo{journal}{International Journal of Solids and Structures}
  \bibinfo{volume}{50}, \bibinfo{pages}{2943--2961}.
\bibitem[{Fickett and Davis(2000)}]{fickett_detonation_2000}
\bibinfo{author}{Fickett, W.}, \bibinfo{author}{Davis, W.C.},
  \bibinfo{year}{2000}.
\newblock \bibinfo{title}{Detonation: theory and experiment}.
\newblock \bibinfo{publisher}{Courier Corporation}.
\bibitem[{Friedlander(1946)}]{friedlander_diffraction_1946}
\bibinfo{author}{Friedlander, F.G.}, \bibinfo{year}{1946}.
\newblock \bibinfo{title}{The diffraction of sound pulses {I}. {Diffraction} by
  a semi-infinite plane}.
\newblock \bibinfo{journal}{Proceedings of the Royal Society of London. Series
  A. Mathematical and Physical Sciences} \bibinfo{volume}{186},
  \bibinfo{pages}{322--344}.
\bibitem[{Islam et~al.(2023a)Islam, Ma, Michopoulos and
  Wang}]{islam_fluid_2023}
\bibinfo{author}{Islam, S.T.}, \bibinfo{author}{Ma, W.},
  \bibinfo{author}{Michopoulos, J.G.}, \bibinfo{author}{Wang, K.},
  \bibinfo{year}{2023}a.
\newblock \bibinfo{title}{Fluid-solid coupled simulation of hypervelocity
  impact and plasma formation}.
\newblock \bibinfo{journal}{International Journal of Impact Engineering} ,
  \bibinfo{pages}{104695}.
\bibitem[{Islam et~al.(2023b)Islam, Ma, Michopoulos and
  Wang}]{islam_plasma_2023}
\bibinfo{author}{Islam, S.T.}, \bibinfo{author}{Ma, W.},
  \bibinfo{author}{Michopoulos, J.G.}, \bibinfo{author}{Wang, K.},
  \bibinfo{year}{2023}b.
\newblock \bibinfo{title}{Plasma formation in ambient fluid from hypervelocity
  impacts}.
\newblock \bibinfo{journal}{Extreme Mechanics Letters} \bibinfo{volume}{58},
  \bibinfo{pages}{101927}.
\bibitem[{Kinney and Graham(2013)}]{kinney_explosive_2013}
\bibinfo{author}{Kinney, G.F.}, \bibinfo{author}{Graham, K.J.},
  \bibinfo{year}{2013}.
\newblock \bibinfo{title}{Explosive shocks in air}.
\newblock \bibinfo{publisher}{Springer Science \& Business Media}.
\bibitem[{Langdon et~al.(2014)Langdon, Ozinsky and
  Yuen}]{langdon_response_2014}
\bibinfo{author}{Langdon, G.S.}, \bibinfo{author}{Ozinsky, A.},
  \bibinfo{author}{Yuen, S.C.K.}, \bibinfo{year}{2014}.
\newblock \bibinfo{title}{The response of partially confined right circular
  stainless steel cylinders to internal air-blast loading}.
\newblock \bibinfo{journal}{International journal of impact engineering}
  \bibinfo{volume}{73}, \bibinfo{pages}{1--14}.
\bibitem[{Li et~al.(2021)Li, Qin and Zhang}]{li_internal_2021}
\bibinfo{author}{Li, J.}, \bibinfo{author}{Qin, Q.}, \bibinfo{author}{Zhang,
  J.}, \bibinfo{year}{2021}.
\newblock \bibinfo{title}{Internal blast resistance of sandwich cylinder with
  lattice cores}.
\newblock \bibinfo{journal}{International Journal of Mechanical Sciences}
  \bibinfo{volume}{191}, \bibinfo{pages}{106107}.
\bibitem[{Liu et~al.(2020)Liu, Gu, Liu, Xu, Hu and Hang}]{liu_dynamic_2020}
\bibinfo{author}{Liu, X.}, \bibinfo{author}{Gu, W.B.}, \bibinfo{author}{Liu,
  J.Q.}, \bibinfo{author}{Xu, J.L.}, \bibinfo{author}{Hu, Y.H.},
  \bibinfo{author}{Hang, Y.M.}, \bibinfo{year}{2020}.
\newblock \bibinfo{title}{Dynamic response of cylindrical explosion containment
  vessels subjected to internal blast loading}.
\newblock \bibinfo{journal}{International journal of impact engineering}
  \bibinfo{volume}{135}, \bibinfo{pages}{103389}.
\bibitem[{M2C(2023)}]{m2c}
\bibinfo{author}{M2C}, \bibinfo{year}{2023}.
\newblock \URLprefix \url{https://github.com/kevinwgy/m2c}.
\bibitem[{Ma et~al.(2010)Ma, Hu, Zheng, Deng and Chen}]{ma_failure_2010}
\bibinfo{author}{Ma, L.}, \bibinfo{author}{Hu, Y.}, \bibinfo{author}{Zheng,
  J.}, \bibinfo{author}{Deng, G.}, \bibinfo{author}{Chen, Y.},
  \bibinfo{year}{2010}.
\newblock \bibinfo{title}{Failure analysis for cylindrical explosion
  containment vessels}.
\newblock \bibinfo{journal}{Engineering Failure Analysis} \bibinfo{volume}{17},
  \bibinfo{pages}{1221--1229}.
\bibitem[{Ma et~al.(2013)Ma, Xin, Hu and Zheng}]{ma_ductile_2013}
\bibinfo{author}{Ma, L.}, \bibinfo{author}{Xin, J.}, \bibinfo{author}{Hu, Y.},
  \bibinfo{author}{Zheng, J.}, \bibinfo{year}{2013}.
\newblock \bibinfo{title}{Ductile and brittle failure assessment of containment
  vessels subjected to internal blast loading}.
\newblock \bibinfo{journal}{International journal of impact engineering}
  \bibinfo{volume}{52}, \bibinfo{pages}{28--36}.
\bibitem[{Ma et~al.(2022)Ma, Zhao, Gilbert and Wang}]{ma_computational_2022}
\bibinfo{author}{Ma, W.}, \bibinfo{author}{Zhao, X.}, \bibinfo{author}{Gilbert,
  C.}, \bibinfo{author}{Wang, K.}, \bibinfo{year}{2022}.
\newblock \bibinfo{title}{Computational analysis of bubble–structure
  interactions in near-field underwater explosion}.
\newblock \bibinfo{journal}{International Journal of Solids and Structures}
  \bibinfo{volume}{242}, \bibinfo{pages}{111527}.
\bibitem[{Main et~al.(2017)Main, Zeng, Avery and Farhat}]{main_enhanced_2017}
\bibinfo{author}{Main, A.}, \bibinfo{author}{Zeng, X.}, \bibinfo{author}{Avery,
  P.}, \bibinfo{author}{Farhat, C.}, \bibinfo{year}{2017}.
\newblock \bibinfo{title}{An enhanced {FIVER} method for multi-material flow
  problems with second-order convergence rate}.
\newblock \bibinfo{journal}{Journal of Computational Physics}
  \bibinfo{volume}{329}, \bibinfo{pages}{141--172}.
\newblock \DOIprefix\doi{10.1016/j.jcp.2016.10.028}.
\bibitem[{Main(2014)}]{main_implicit_2014}
\bibinfo{author}{Main, G.A.}, \bibinfo{year}{2014}.
\newblock \bibinfo{title}{Implicit and higher-order discretization methods for
  compressible multi-phase fluid and fluid-structure problems}.
\newblock \bibinfo{publisher}{Stanford University}.
\bibitem[{Pickerd et~al.(2016)Pickerd, Bornstein, McCarthy and
  Buckland}]{pickerd_analysis_2016}
\bibinfo{author}{Pickerd, V.}, \bibinfo{author}{Bornstein, H.},
  \bibinfo{author}{McCarthy, P.}, \bibinfo{author}{Buckland, M.},
  \bibinfo{year}{2016}.
\newblock \bibinfo{title}{Analysis of the structural response and failure of
  containers subjected to internal blast loading}.
\newblock \bibinfo{journal}{International journal of impact engineering}
  \bibinfo{volume}{95}, \bibinfo{pages}{40--53}.
\bibitem[{Rokhy and Mostofi(2023)}]{rokhy_tracking_2023}
\bibinfo{author}{Rokhy, H.}, \bibinfo{author}{Mostofi, T.M.},
  \bibinfo{year}{2023}.
\newblock \bibinfo{title}{Tracking the explosion characteristics of the
  hydrogen-air mixture near a concrete barrier wall using cese ibm fsi solver
  in ls-dyna incorporating the reduced chemical kinetic model}.
\newblock \bibinfo{journal}{International Journal of Impact Engineering}
  \bibinfo{volume}{172}, \bibinfo{pages}{104401}.
\bibitem[{Rokhy and Soury(2022)}]{rokhy_investigation_2022}
\bibinfo{author}{Rokhy, H.}, \bibinfo{author}{Soury, H.}, \bibinfo{year}{2022}.
\newblock \bibinfo{title}{Investigation of the confinement effects on the blast
  wave propagated from gas mixture detonation utilizing the cese method with
  finite rate chemistry model}.
\newblock \bibinfo{journal}{Combustion Science and Technology}
  \bibinfo{volume}{194}, \bibinfo{pages}{3003--3020}.
\bibitem[{Simo and Hughes(2006)}]{simo_computational_2006}
\bibinfo{author}{Simo, J.}, \bibinfo{author}{Hughes, T.}, \bibinfo{year}{2006}.
\newblock \bibinfo{title}{Computational Inelasticity}.
\newblock \bibinfo{publisher}{Springer Science \& Business Media}.
\bibitem[{Souli et~al.(2000)Souli, Ouahsine and Lewin}]{souli_ale_2000}
\bibinfo{author}{Souli, M.}, \bibinfo{author}{Ouahsine, A.},
  \bibinfo{author}{Lewin, L.}, \bibinfo{year}{2000}.
\newblock \bibinfo{title}{{ALE} formulation for fluid–structure interaction
  problems}.
\newblock \bibinfo{journal}{Computer Methods in Applied Mechanics and
  Engineering} \bibinfo{volume}{190}, \bibinfo{pages}{659--675}.
\bibitem[{Taylor(1950)}]{taylor_dynamics_1950}
\bibinfo{author}{Taylor, G.I.}, \bibinfo{year}{1950}.
\newblock \bibinfo{title}{The dynamics of the combustion products behind plane
  and spherical detonation fronts in explosives}.
\newblock \bibinfo{journal}{Proceedings of the Royal Society of London. Series
  A. Mathematical and Physical Sciences} \bibinfo{volume}{200},
  \bibinfo{pages}{235--247}.
\bibitem[{Wang et~al.(2012)Wang, Grétarsson, Main and
  Farhat}]{wang_computational_2012}
\bibinfo{author}{Wang, K.}, \bibinfo{author}{Grétarsson, J.},
  \bibinfo{author}{Main, A.}, \bibinfo{author}{Farhat, C.},
  \bibinfo{year}{2012}.
\newblock \bibinfo{title}{Computational algorithms for tracking dynamic
  fluid-structure interfaces in embedded boundary methods}.
\newblock \bibinfo{journal}{International Journal for Numerical Methods in
  Fluids} \bibinfo{volume}{70}, \bibinfo{pages}{515--535}.
\newblock \DOIprefix\doi{10.1002/fld.3659}.
\bibitem[{Wang(2017)}]{wang_multiphase_2017}
\bibinfo{author}{Wang, K.G.}, \bibinfo{year}{2017}.
\newblock \bibinfo{title}{Multiphase fluid-solid coupled analysis of
  shock-bubble-stone interaction in shockwave lithotripsy}.
\newblock \bibinfo{journal}{International journal for numerical methods in
  biomedical engineering} \bibinfo{volume}{33}, \bibinfo{pages}{e2855}.
\bibitem[{Wang et~al.(2015)Wang, Lea and Farhat}]{wang_computational_2015}
\bibinfo{author}{Wang, K.G.}, \bibinfo{author}{Lea, P.},
  \bibinfo{author}{Farhat, C.}, \bibinfo{year}{2015}.
\newblock \bibinfo{title}{A computational framework for the simulation of
  high‐speed multi‐material fluid–structure interaction problems with
  dynamic fracture}.
\newblock \bibinfo{journal}{International Journal for Numerical Methods in
  Engineering} \bibinfo{volume}{104}, \bibinfo{pages}{585--623}.
\bibitem[{Zhao et~al.(2023a)Zhao, Ma, Chen, Xiang, Zhong and
  Wang}]{zhao_long_2023}
\bibinfo{author}{Zhao, X.}, \bibinfo{author}{Ma, W.}, \bibinfo{author}{Chen,
  J.}, \bibinfo{author}{Xiang, G.}, \bibinfo{author}{Zhong, P.},
  \bibinfo{author}{Wang, K.}, \bibinfo{year}{2023}a.
\newblock \bibinfo{title}{Long-pulse laser-induced cavitation: A race between
  advection and phase transition}.
\newblock \bibinfo{journal}{arXiv preprint arXiv:2308.11866} .
\bibitem[{Zhao et~al.(2023b)Zhao, Ma and Wang}]{zhao_simulating_2023}
\bibinfo{author}{Zhao, X.}, \bibinfo{author}{Ma, W.}, \bibinfo{author}{Wang,
  K.}, \bibinfo{year}{2023}b.
\newblock \bibinfo{title}{Simulating laser-fluid coupling and laser-induced
  cavitation using embedded boundary and level set methods}.
\newblock \bibinfo{journal}{Journal of Computational Physics}
  \bibinfo{volume}{472}, \bibinfo{pages}{111656}.

\end{thebibliography}
	
	\bio{}
	\endbio
	
	\endbio
	
\end{document}